\documentclass[prd,aps,floats,floatfix,superscriptaddress,preprintnumbers,showpacs,eqsecnum,nofootinbib,twocolumn]{revtex4}
\usepackage{amsfonts,amsmath,epsfig,amssymb,latexsym,color,graphicx,array,bm,float }

\definecolor{red}{rgb}{1,0,0}
\definecolor{blue }{rgb}{0,0,1}
\definecolor{green}{rgb}{0,1,0}

\newcommand{\bea}{\begin{eqnarray}}
\newcommand{\ena}{\end{eqnarray}}

\newcommand{\hs}[1]{\hspace{#1 mm}}
\renewcommand{\a}{\alpha}
\renewcommand{\b}{\beta}
\renewcommand{\c}{\gamma}

\newcommand{\dsl}{\pa \kern-0.5em /}
\newcommand{\la}{\lambda}
\newcommand{\pa}{\partial}

\newcommand{\nn}{\nonumber\\}
\newcommand{\p}[1]{(\ref{#1})}

\newcommand{\vect}[1]{\!\!\!\mbox{ \boldmath $#1$}}

\begin{document}

\preprint{WU-AP/312/11}
\preprint{KU-TP 053~~~~~\,}

\title{Accelerating Universes in String Theory
via Field Redefinition}

\author{Kei-ichi Maeda}
\email{maeda"at"waseda.jp}
\affiliation{
Department of Physics, Waseda University,
Shinjuku, Tokyo 169-8555, Japan}
\affiliation{
Research Institute for Science and Engineering, Waseda University,
Shinjuku, Tokyo 169-8555, Japan}

\author{Nobuyoshi Ohta}
\email{ohtan"at"phys.kindai.ac.jp}
\affiliation{
Department of Physics, Kinki University,
Higashi-Osaka, Osaka 577-8502, Japan}

\author{Ryo Wakebe}
\email{wakebe"at"gravity.phys.waseda.ac.jp}
\affiliation{
Department of Physics, Waseda University,
Shinjuku, Tokyo 169-8555, Japan}

\begin{abstract}
We study cosmological solutions in the effective heterotic string theory with
$\alpha'$-correction terms in string frame.
It is pointed out that the effective theory has an ambiguity via field redefinition
and we analyze generalized effective theories due to this ambiguity.
We restrict our analysis to the effective theories which give equations of motion
of second order in the derivatives, just as ``Galileon" field theory.
This class of effective actions contains two free coupling constants.
We find de Sitter solutions as well as the power-law expanding universes in
our four-dimensional Einstein frame.
The accelerated expanding universes are always the attractors
in the present dynamical system.
\end{abstract}

\maketitle



\section{Introduction}

The recent cosmological observations have confirmed the accelerated
expansion of the present universe~\cite{wmap}.
We also know that an inflationary cosmological epoch may exist
in the early stage of the universe
\cite{Starobinsky,inflation1,inflation2,inflation3,inflation4}.
It is possible to construct cosmological
models with such accelerating phases if one introduces a new scalar field
with an appropriate potential. However, it is desirable to derive a natural
model from a fundamental theory of particle
physics without introducing any unknown field. The most promising candidate
for such a fundamental theory
is the ten-dimensional superstring theory or eleven-dimensional M theory.
They are hoped to give an interesting explanation
for the accelerated expansion of the
universe upon compactification to four dimensions.

In the low-energy effective field theories of superstrings or supergravity,
there is a no-go theorem, however, which forbids such
accelerated expanding spacetime solutions if six- or
seven-dimensional internal space is a time-independent nonsingular compact
manifold without boundary~\cite{no-go}. This theorem also assumes that
the gravitational action is given only
by the Einstein-Hilbert curvature term. So in order to  evade this theorem,
we have to violate some of those assumptions.
In fact, it has been shown that some model
with a certain period of the accelerated expansion is obtained from the
higher-dimensional vacuum Einstein equation if one assumes a time-dependent
hyperbolic internal space~\cite{TW}. It has been shown~\cite{NO} that this
class of models is obtained from what are known as
S-branes~\cite{Wohlfarth:2003ni,Sbrane1,Sbrane2}
in the limit of vanishing flux of three-form fields (see also~\cite{Sbrane3}).
For other attempts for inflation in the context of string theory
by use of a time-dependent internal space,
see, for instance, Refs.~\cite{other,cosm2,cosm3}.
Unfortunately, this class
of models does not give a sufficiently long rapid expansion phase to
resolve the cosmological problems.

If one introduces branes, it also violates the assumptions in the no-go theorem.
There are many discussions about this type of brane inflations~\cite{brane_inflation,Brane_rev}.
Although some of them could be promising,
they still require a fine-tuning in their setup.
We also do not know which model is most favorite one.

The final possible violation of the assumptions in the no-go theorem
is a change of gravitational action.
The inflation based on the fundamental theory
is expected to occur at the Planck scale in
ten or eleven dimensions. With phenomena at such a high-energy scale,
 we cannot ignore quantum corrections in the effective action,
at least in the early stage of the universe.
The higher-order correction terms in the curvatures
appear via quantum effects in the field-theory limit of
the superstring theory or M theory~\cite{MT,hetero0,hetero,Mth,TBB}.
Those terms should be included to the lowest supergravity action.
The no-go theorem no longer applies to such theories with higher curvature terms.
With those corrections, the inflationary scenario will be significantly
affected, at least in the early stage of the universe.

In fact, Starobinsky first presented an inflationary scenario in
the effective theory with the higher curvature correction terms,
which are given by quantum one-loop correction
of fundamental fields~\cite{Starobinsky,maeda1}.
Furthermore, if one believes superstring theory, or supergravity
theories, which are effective ones in the field-theory limit of
superstring theory, we have to study cosmology in ten dimensions.
The cosmological models in higher dimensions were studied
as Kaluza-Klein cosmology intensively in
the 80's by many authors~\cite{KK_cosmology,maeda}.
The effect of higher curvature correction terms
was also analyzed~\cite{ISHI,old1,MO,BGO}.

The leading term of quadratic curvature corrections for the heterotic string theory
is the Gauss-Bonnet (GB) combination~\cite{MT,hetero0,hetero}.
This model without a dilaton was studied in Ref.~\cite{ISHI} and it was shown that
there are two solutions, in which three-dimensional space is exponentially expanding.
One may call them generalized de Sitter solutions since
the internal space is also expanding or contracting exponentially.

However, it does not mean that such solutions provide us a successful
inflationary scenario, for which we have several observational constraints.
The period of inflation must be longer than 50-60 e-folding time, and
inflation must end, followed by reheating of the universe.
The density fluctuation should be obtained with an appropriate value.
Most of inflationary scenarios have been analyzed based on 4-dimensional
Einstein gravity. Hence if we work in the 4-dimensional Einstein gravity
(plus some correction terms), it would be easier to check whether such a model gives
a successful inflationary scenario or not.
In higher-dimensional gravitational theories,
we can extract four-dimensional
spacetime to find the Einstein-Hilbert action (plus some correction terms)
in four dimensions, which we call the 4D Einstein frame.
In such a 4D Einstein frame, the gravitational constant becomes really constant. It makes easy to discuss cosmology in four dimensions because we can adopt
a lot of conventional approaches and results.
On the other hand, if one works in 10-dimensional spacetime,
we have to analyze the model much more carefully.
The volume moduli must be fixed at present in order to avoid
a harmful massless scalar field in four dimensions, while it may be free in
the early stage. As a result, such a cosmology  may depend on the detail of
a moduli fixing mechanism, which we do not know.
We thus conclude that it is better to analyze a model in the 4D Einstein frame,
which will provide us a sufficient condition for a successful scenario.

As for the above generalized de Sitter solutions, we find
our three-dimensional universe does not expand
exponentially in the 4D Einstein frame.
Instead, one of the solutions shows a decelerating expansion with positive power
exponent less than unity.
In the other solution with negative power exponent, the scale factor $a$ of
our universe diverges as $t\rightarrow 0$. It is called
a pole-type inflation found in Kaluza-Klein cosmology~\cite{Sahdev}.
In this limit we find an accelerating expansion
($a\rightarrow \infty$ as $t\rightarrow 0$).
Here, however, the effective field theory may be no longer valid
because it evolves into a stringy state where the internal space gets very small.
Since we do not know what happens in such a stringy state,
one may not conclude that an inflation
(rapid expansion) can be realized with the higher curvature terms.

We also have to include the effect of a dilaton field, unless we show its
stabilization mechanism. Many studies of higher curvature corrections
consider a pure GB term without a dilaton field, or assume that the dilaton field
is constant, which may not be consistent with the full equations of motion
in the effective string theory. We have to include the dynamics of the dilaton field.

The effective action in the heterotic string theory has been calculated~\cite{MT}.
The authors discuss a field theory action generating the S-matrix
which coincides with the massless sector of the (tree-level)
string S-matrix. The generating functional for
the string S-matrix is represented as a path integral over surfaces with the free
string action replaced by the generalized $\sigma$-model action describing
the string propagating in a non-trivial background.
They found the Riemann curvature squared term as
$\alpha'$-corrections~\cite{MT}.
Using the freedom of field redefinition, the theory is transformed into the GB
combination together with higher-order derivative terms of the dilaton field.

In Ref.~\cite{BGO}, the authors have analyzed the possibility of inflation
in the Einstein-Gauss-Bonnet-dilaton system in the 10-dimensional
Einstein frame.
There the considered system contains only the GB term as the quantum corrections
as well as the Einstein-Hilbert curvature term and canonical kinetic term
of the dilaton field. The  higher-order derivative terms of the dilaton field
are ignored. They have shown that there are several fixed points, some of which
give generalized de Sitter solutions, i.e., our three-dimensional space is
exponentially expanding and the internal space is
also exponentially changing just as the case without a dilaton.
In this case, the dilaton is also time-dependent.
In the four-dimensional Einstein frame, which describes
gravity in our world, either spacetime does not show
any accelerating expansion, or it expands with negative power exponent.
Although the latter case gives the accelerating expansion, it goes into a stringy
state. We may not conclude in this field-theoretical approach that inflation can
be realized with such higher curvature terms, just as the case without a dilaton.
Related discussions of impossibility of de Sitter solutions are given in~\cite{GMQS}.
Density perturbations are also studied for such theories~\cite{GOT,GS}.

As in the above study, when we discuss cosmological solutions, often only the
GB term is taken into account as quantum corrections and the higher-derivative
terms of the dilaton field is ignored. However, it is known that the effective theory
does contain such higher-order derivative terms of the dilaton~\cite{MT},
and we have to consider these higher order terms in general.

A related issue is the difference of the frames.
In the string frame, the Einstein-Hilbert curvature term is coupled to the dilaton
field whereas in the Einstein frame it is not.
In order to find the theory in one of the frames from the other, we perform a conformal
transformation by the dilaton field. We then find additional higher-derivative terms of
the dilaton as well as their coupling to the curvatures from the GB term~\cite{conf,footnote1}
but this effect can be incorporated by including these terms.

In the present order of quantum corrections,
there is also more important ambiguity in the effective action coming from the field
redefinition. Since the perturbative S-matrix does not change under the local field
redefinitions, we have a freedom from this in the effective action.
This produces a difference in the higher-order derivative terms of the dilaton field and
the curvatures when further higher order terms, which are not known, are ignored.
Taking such a field redefinition ambiguity into account, we are lead to a large class
of effective actions, all of which correspond to the same string S-matrix.
Of course, when we include full orders of quantum corrections, the obtained
effective action must be unique and there cannot be any physically new solutions that
were not present in the solution space prior to field redefinitions.
Unfortunately, we do not know how to find such a self-consistent full-order
effective action. As long as we try to obtain the effective action perturbatively
in $\a'$, there is always ambiguity due to the field redefinition which produces
terms higher order than already known. If all these terms are included,
we should get the same solutions as before, but since we expect that there are other
higher order terms, it is reasonable to try to find solutions without these
higher order terms. Certainly, if we expand such an exact effective action, if any,
we will find the correct effective action in the present order of quantum corrections.
We can expect that such theory falls into the class of theories that we are considering.
It is thus significant to study cosmology and the possibility of inflation
in this generalized class of effective theories obtained by field redefinition.

In this paper, we study the effect on cosmology of such ambiguity
in the field redefinition in the string frame.
The action we analyze is obtained via field redefinition from the effective
action with $\alpha'$-correction terms of the heterotic string theory.
We restrict our analysis to the case such that the equations of motion contain
up to the second-order derivatives just as the ''Galileon" theory.
The curvature squared term is given by the GB combination,
while the other higher-order correction terms to the dilaton are given by
two free coupling constants.

To analyze the present model, we adopt the method of dynamical system.
We reduce the basic equations into an autonomous system, and find the fixed points and
analyze their stabilities.
A similar approach was used to study inflationary solutions in M
theory with fourth-order quantum corrections~\cite{MO}.

This paper is organized as follows. In Sec.~II, we first
give the effective action which we will discuss.
In Sec.~III,  we present the explicit forms of the basic equations for
higher-dimensional cosmology assuming an appropriate metric form.
We show that those equations form an autonomous system.
In Sec.~IV, we look for the fixed points, which correspond
to the accelerating universes in four-dimensional
Einstein frame, and  analyze their stabilities.
We find that de Sitter expanding
universe is possible in our four-dimensional Einstein frame
and it is a stable attractor in the present dynamical system.
Sec.~V is devoted to conclusion and discussion.

In Appendix A,
we present the fixed points explicitly for some special cases.
In order to see the frame dependence,
we also discuss cosmological solutions in the Einstein-Gauss-Bonnet-dilaton
 system in the string frame in Appendix B.
Since the similar setup in the Einstein frame was analyzed in Ref.~\cite{BGO},
we compare our results with theirs.
We find that there is not much difference in the cosmological solutions.

\begin{widetext}
\vskip 5mm

\section{Effective Action and Field Redefinition}
\label{effective_action}

We start with the effective action,
which describes the low-energy dynamics of the massless string modes.
With the $\alpha'$ correction in the effective action~\cite{MT}, we have, in one
scheme,
\bea
S = \frac{1}{2\kappa_D^2} \int d^D x \sqrt{-g} e^{-2\phi}
\Big[ R+4 (\nabla \phi)^2
+ \alpha_2 R_{ABCD}^2
\Big]
\,,
\label{ea_Tseytlin}
\ena
where $\phi$ is a dilaton field,
$\kappa_D^2$ is a $D$-dimensional gravitational constant,
and $\alpha_2={\alpha'/8}$ is a coupling constant to the curvature square term
with the Regge slope parameter $\a'$.
We drop the contributions from the  NS-NS forms and fermions.
Since this correction is obtained from the string $S$-matrix,
there is an ambiguity in the effective action caused by
the field redefinition $g_{AB}\rightarrow  g_{AB}+ \delta g_{AB} $
and $ \phi \rightarrow \phi + \delta \phi $, where
\bea
\delta g_{AB} &=& \a_2\left\{ b_1 R_{AB}+b_2 \nabla_A \phi
 \nabla_B \phi
+g_{AB}[b_3 R+b_4(\nabla\phi)^2+b_5 \nabla^2 \phi]\right\}, \nn
\delta \phi &=& \a_2 \left\{ c_1 R+ c_2(\nabla \phi)^2+c_3 \nabla^2 \phi \right\},
\label{field_redef}
\ena
with $b_i$'s and $c_i$'s being arbitrary constants.
This field redefinition gives the general effective action up to $O(\a')$ as
\bea
S
&=& \frac{1}{2\kappa_D^2} \int d^D x \sqrt{-g} e^{-2\phi}
\Big\{  R+4 (\nabla \phi)^2 + \a_2 \Big[R_{ABCD}^2
+ b_1 R_{AB}^2
+(b_2+4 b_1) R_{AB} \nabla^A \phi \nabla^B \phi
\nn
&&
+\Big(2c_1-\frac{1}{2}b_1-\frac{D-2}{2}b_3\Big)R^2
+\Big(2c_2-8c_1-\frac{b_2}{2}+2Db_3-\frac{D-2}{2}b_4 \Big) R(\nabla\phi)^2 \nn
&& +\Big(2c_3+8c_1-b_1-2(D-1)b_3-\frac{D-2}{2}b_5 \Big) R (\nabla^2\phi)
-2\left(4c_2- 2b_2-Db_4\right) (\nabla\phi)^4 \nn
&& +[8c_2-8c_3-3b_2-2(D-1)b_4+2Db_5] \Box\phi(\nabla\phi)^2
+[8c_3-2(D-1)b_5] (\Box\phi)^2 \Big]
\Big\}
\,.
\ena
It is certainly true that if we keep $O(\a'^2)$ or higher-order terms in the result,
we may get the equivalent theory.
But they are not physically relevant and it does not make much sense to discuss their
effects because the effective theory to that order is not known or ignored
already in \p{ea_Tseytlin}.
Thus there is an intrinsic ambiguity in the effective theory.
However we may restrict the theory to certain extent by imposing some consistency conditions.

In our approach, we may take the viewpoint that the correct effective theory
must not have any ghost or tachyon.
The curvature square term may be given by the GB combination,
$R_{\rm GB}^2\equiv R_{ABCD}^2-4R_{AB}^2+R^2$,
which gives the second-order differential equations.
Hence it is likely that the correct effective theory is the one
which gives equations of motion without derivatives of order higher than two,
just like the ``Galileon'' theory~\cite{galileon}.
Here we take such an effective action and require that higher derivative
terms do not appear in the resulting field equations. This condition
gives the constraints on the coefficients $b_i$'s and $c_i$'s
in the field redefinition (\ref{field_redef}):
\bea
&& b_1=-4,~~
\nn
&&
2c_1-\frac{1}{2}b_1-\frac{D-2}{2}b_3=1,
\nn
&& 2c_2-8c_1-\frac{b_2}{2}+2Db_3-\frac{D-2}{2}b_4 =-{1\over 2}
\Big(b_2+4 b_1\Big),
\nn
&&
2c_3+8c_1-b_1-2(D-1)b_3-\frac{D-2}{2}b_5 =0,
\nn
&&
8c_3-2(D-1)b_5=0
\,,
\ena
which are solved by
\bea
&& b_1=-4,~~
b_5=4 b_3,~~\nn
&& c_1=\frac{D-2}{4} b_3-{1\over 2},~~
c_2=-2 b_3+\frac{D-2}{4}b_4+2,~~
c_3=(D-1)b_3\,,
\ena
where $b_2, b_3,$ and $b_4$ are free.
However the coefficients of non-trivial terms in the effective action
are not independent. As a result, we find
 a two-parameter family of the effective theory:
\bea
S &=& \frac{1}{2\kappa_D^2} \int d^D x \sqrt{-g} e^{-2\phi}
\Big[ R+4 (\nabla \phi)^2
 \nn
&& \hs{10}
+ \a_2 \Big\{R_{GB}^2 +\lambda (\nabla \phi)^4
 +\mu \Big(R^{AB}-\frac12 Rg^{AB}\Big)
\nabla_A\phi\nabla_B\phi+\nu  \Box\phi(\nabla\phi)^2 \Big\} \Big]\,,
\label{gali}
\ena
with $\lambda+2(\mu+\nu)+16=0$. $\mu$ and $\nu$ are two free parameters.
We will analyze the cosmological solutions in this class of effective theories.
Note that the effective theory only with the GB term for
$\alpha'$-corrections in the string frame as well as that
in the Einstein frame considered in \cite{BGO} are not involved in this family,
since in that case $\la=\mu=\nu=0$.
(See \cite{BGO} and Appendix \ref{append2} for cosmological solutions for
such an effective theory.)

\section{basic equations}

Using the class of effective theories with $\alpha'$-corrections given
in Sec.~\ref{effective_action}, we discuss cosmological solutions
and their properties.
Let us assume the following metric form in $D$-dimensional space:
\bea
ds_D^2=-e^{2u_0(t)}dt^2 + e^{2u_1(t)}ds_p^2 + e^{2u_2(t)}ds_q^2 \,,
\label{metric}
\ena
where $D=1+p+q$. Both the $p$-dimensional space ($ds_p^2$)
and $q$-dimensional one ($ds_q^2$) are
chosen to be maximally symmetric, with the
curvature signatures  of $\sigma_p (=0,\pm 1)$ and
 $\sigma_q (=0,\pm 1)$, respectively.

With the above metric ansatz, we can simplify the action as
\bea
S={1\over 2\kappa_D^2}V_pV_q\int dt\left[
{\cal L}_{\bf 0}+ \alpha_2\left({\cal L}_{\rm I}+ {\cal L}_{\rm II}\right) \right]
\,,
\label{EA}
\ena
where ${\cal L}_{\bf 0}$,
${\cal L}_{\rm I}$,
and ${\cal L}_{\rm II}$ are the lowest Lagrangian
from the Einstein-Hilbert action and the canonical kinetic term of
the dilaton field,
that from the GB term, and that from the other
$\alpha'$-correction terms, respectively,
with
\bea
{\cal L}_{\bf 0}
&=& e^{-u_0+pu_1+qu_2-2\phi}\Bigl[p_1A_p + q_1A_q - 2(p_1\dot u_1{^2} +
pq\dot u_1\dot u_2 + q_1\dot u_2{^2} ) +4 (p\dot u_1 + q\dot u_2) \dot\phi
- 4 \dot\phi^2 \Bigr]\,,~~
\label{EA0}
\\
{\cal L}_{\rm I} &=& {1\over 3} e^{-3u_0+pu_1+qu_2-2\phi}\Bigl\{3\left(
 p_3A_p{^2}
+ 2p_1q_1A_qA_p+ q_3A_q{^2}\right) \nn
&& -12\left[
A_p (p_3\dot u_1{^2}+p_2q\dot u_1\dot u_2 + p_1q_1\dot u_2{^2})
+A_q(p_1q_1\dot u_1{^2}+pq_2\dot u_1\dot u_2 + q_3\dot u_2{^2})
\right]
\nn
&& + 4\left(2p_3\dot u_1{^4} + 2p_2q \dot u_1{^3}\dot u_2
+ 3p_1q_1\dot u_1{^2}\dot u_2{^2} +2pq_2\dot u_1\dot u_2{^3}
+ 2 q_3\dot u_2{^4}\right)
\nn
&& + 8 \dot \phi \Big[ 3(p_2 \dot u_1+p_1 q \dot u_2)A_p
+3 (pq_1 \dot u_1+ q_2 \dot u_2)A_q - 2 \Bigl( p_2 \dot u_1^3 + q_2 \dot u_2^3
\Bigr)\Big]\Bigr\}\,,
\label{EAGB}
\\
{\cal L}_{\rm II} &=&{1\over 6}  e^{-3u_0+pu_1+qu_2-2\phi} \dot{\phi}^2\Big[ 3
 \mu( p_1 A_p+q_1A_q+2 p q \dot{u}_1 \dot{u}_2)
+2\left(2\nu +3\lambda \right)\dot{\phi}^2
+4\nu\dot{\phi}(p\dot{u}_1+q\dot{u}_2)\Big]
\label{EAphi}
\,,
~~~~~
\ena
and we drop the surface terms.
$V_p$ and $V_q$ are the volumes of the $p$-space $ds_p^2$
and the $q$-space $ds_q^2$, respectively, and
the following notations are introduced:
\bea
&& A_p \equiv \dot{u}_1^2+\sigma_p e^{2(u_0-u_1)}\,,\quad
A_q \equiv \dot{u}_2^2+\sigma_q e^{2(u_0-u_2)}\,,
\label{Apq}
\\
&& (\ell -m)_n \equiv  (\ell -m)(\ell -m-1)(\ell -m-2)\cdots (\ell -n) \,,
\label{pqmn}
\ena
with $\ell, m,n$ being positive integers ($\ell>n>m$).
Taking the variation with respect to $u_0,u_1,u_2$, and $\phi$,
we obtain four field equations:
\bea
\label{rfe1}
{\cal F}~~ &\equiv& {\cal F}_{\bf 0} +\a_2( {\cal F}_{\rm I}
 +{\cal F}_{\rm II}) =0\,, \\
\label{rfe2}
{\cal F}^{(p)} &\equiv& f_{\bf 0}^{(p)} + \a_2( f_{\rm I}^{(p)} +f_{\rm II}^{(p)})
+ X \left[g_{\bf 0}^{(p)} + \a_2\left( g_{\rm I}^{(p)}
+g_{\rm II}^{(p)}\right)\right]\nn
&&+ Y\left[h_{\bf 0}^{(p)} + \a_2\left( h_{\rm I}^{(p)}
+h_{\rm II}^{(p)}\right)\right]
 - Z \left[i_{\bf 0}^{(p)}+\a_2\left(i_{\rm I}^{(p)}
+i_{\rm II}^{(p)}\right)\right]=0\,, \\
\label{rfe3}
{\cal F}^{(q)} &\equiv& f_{\bf 0}^{(q)} +\a_2(  f_{\rm I}^{(q)}
+f_{\rm II}^{(q)}) + Y \left[g_{\bf 0}^{(q)}
+ \a_2\left( g_{\rm I}^{(q)}+g_{\rm II}^{(q)}\right)\right]
\nn
&& + X\left[h_{\bf 0}^{(q)} + \a_2\left( h_{\rm I}^{(q)}+h_{\rm II}^{(q)}\right)\right]
- Z \left[i_{\bf 0}^{(q)}+\a_2\left( i_{\rm I}^{(q)}
+i_{\rm II}^{(q)}\right)\right] =0\,,\\
\label{rfe4}
{\cal F}^{(\phi)} &\equiv&  f_{\bf 0}^{(\phi)}
+\a_2\left( f_{\rm II}^{(\phi)}+g_{\rm II}^{(\phi)}X
+h_{\rm II}^{(\phi)}Y
+i_{\rm II}^{(\phi)}Z\right)-\frac{\a_2}{4} e^{2u_0} R_{\rm GB}^2 =0,~~~
\ena
where
\bea
X &\equiv& \ddot u_1 - \dot u_0 \dot u_1 +\dot u_1^2\,, \quad
Y  ~\equiv~  \ddot u_2 - \dot u_0 \dot u_2 +\dot u_2^2 \,, \quad
Z ~\equiv~  \ddot\phi -(\dot u_0 - p\dot u_1 - q\dot u_2)\dot\phi\,, \nn
{\cal F}_{\bf 0}&=& p_1 A_p+q_1 A_q+2pq\dot{u}_1\dot{u}_2 -4(p\dot u_1 + q\dot u_2)
\dot\phi + 4 \dot\phi^2\,, \nn
f_{\bf 0}^{(p)}&=& (p-1)_2A_p+q_1 A_q+2(p-1)q\dot{u}_1\dot{u}_2 +4\dot\phi \dot u_1
+4 \dot\phi^2\,, \nn
f_{\bf 0}^{(q)}&=& p_1 A_p+(q-1)_2A_q+2p(q-1)\dot{u}_1\dot{u}_2 +4\dot\phi \dot u_2
+4 \dot\phi^2\,, \nn
f_{\bf 0}^{(\phi)}&=&-\frac{1}{4}\left(p_1A_p+q_1A_q+2pX+2qY-4Z
+4\dot{\phi}^2+2pq\dot{u}_1\dot{u}_2\right)\,,\nn
g_{\bf 0}^{(p)} &=& 2(p-1)\,,~~
g_{\bf 0}^{(q)}= 2(q-1)\,,~~
h_{\bf 0}^{(p)}= 2q\,, ~~
h_{\bf 0}^{(q)}= 2p\,,~~
i_{\bf 0}^{(p)}= i_{\bf 0}^{(q)}= 4 \,,
\label{eh1}
\\[1em]
{\cal F}_{\rm I} &=& e^{-2u_0} \Big\{ p_3 A_p^2+2p_1q_1 A_pA_q
+q_3 A_q^2 + 4(p_2 q A_p+p q_2 A_q + p_1q_1\dot{u}_1 \dot{u}_2)\dot{u}_1\dot{u}_2 \nn
&& 
-\; 8\dot\phi \big[ (p_2 \dot u_1 +p_1 q \dot u_2)A_p
+ (p q_1 \dot u_1 + q_2 \dot u_2)A_q + 2( p_1 q \dot u_1 + pq_1 \dot u_2)
 \dot u_1 \dot u_2 \big] \Big\}, \nn
f_{\rm I}^{(p)}\; &=&\;  e^{-2u_0}\Big\{
(p-1)_4A_p^2+2(p-1)_2q_1 A_pA_q+q_3 A_q^2
+\; 4\left[(p-1)_3qA_p+(p-1)q_2 A_q
\right.
\nn
&& 
\left. +(p-1)_2q_1\dot{u}_1\dot{u}_2
\right]\dot{u}_1\dot{u}_2 +\; 8 \dot \phi \big[ ((p-1)_2 A_p + q_1 A_q
+ 2(p-1)q \dot u_1 \dot u_2 )
(\dot u_1 + 2 \dot\phi)
\nn
&& 
+\; 2 ( (p-1)_2 \dot u_1 A_p + q_1 \dot u_2 A_q +(p-1)q \dot u_1 \dot u_2
(\dot u_1 +\dot u_2) ) \big] \Big\},
\nn
f_{\rm I}^{(q)}\; &=&\;  e^{-2u_0}\Big\{
p_3 A_p^2+ 2p_1 (q-1)_2 A_pA_q+(q-1)_4A_q^2
+\; 4 \left[ p_2 (q-1)A_p+p(q-1)_3A_q
\right.
\nn
&& 
\left.
+p_1(q-1)_2\dot{u}_1 \dot{u}_2
 \right]\dot{u}_1\dot{u}_2
+\; 8  \dot \phi \big[ ( p_1 A_p + (q-1)_2 A_q + 2p(q-1) \dot u_1 \dot u_2 )
 (\dot u_2 + 2 \dot\phi) \nn
&& 
+\; 2 ( p_1 \dot u_1 A_p +(q-1)_2 \dot u_2 A_q + p(q-1) \dot u_1 \dot u_2
(\dot u_1 +\dot u_2) ) \big] \Big\}, \nn
g_{\rm I}^{(p)} &=& 4(p-1) e^{-2u_0} \Big[
(p-2)_3 A_p + q_1 A_q+ 2(p-2)q \dot u_1 \dot u_2
- 4 ((p-2)\dot u_1 + q \dot u_2 ) \dot\phi \Big], \nn
g_{\rm I}^{(q)} &=& 4(q-1) e^{-2u_0} \Big[
p_1 A_p + (q-2)_3 A_q + 2 p (q-2) \dot u_1 \dot u_2
- 4 (p \dot u_1 + (q-2) \dot u_2 ) \dot\phi \Big],
\nonumber
\ena
\bea
&& h_{\rm I}^{(p)}\; =\; 4q e^{-2u_0}\Big[
(p-1)_2A_p+(q-1)_2A_q+2(p-1)(q-1)\dot{u}_1\dot{u}_2
-\; 4 ((p-1)\dot u_1 +(q-1) \dot u_2) \dot \phi\Big], \nn
&& h_{\rm I}^{(q)}\; =\; 4p e^{-2u_0}\Big[
(p-1)_2A_p+(q-1)_2A_q+2(p-1)(q-1)\dot{u}_1\dot{u}_2
-\; 4((p-1)\dot u_1 +(q-1) \dot u_2) \dot \phi\Big],
\nn
&& i_{\rm I}^{(p)}\; =\; 8 e^{-2u_0} \Big[(p-1)_2 A_p
+ q_1 A_q + 2(p-1)q \dot u_1 \dot u_2\Big], \nn
&& i_{\rm I}^{(q)}\; =\; 8 e^{-2u_0}  \Big[p_1 A_p + (q-1)_2 A_q
+ 2p(q-1) \dot u_1 \dot u_2\Big],
\label{gb2}
\ena
\bea
&&
{\cal F}_{\rm II}\; =  -{1\over 2}e^{-2u_0 }
\dot{\phi}^2\Bigl\{\mu \Bigl[ p_1A_p+ q_1A_q +
2 (p_1\dot{u}_1^2+3pq\dot{u}_1\dot{u}_2+q_1\dot{u}_2^2)\Bigr]
+2\dot{\phi}\left[(2\nu +3\lambda)\dot{\phi}+2\nu (p\dot{u}_1
+q\dot{u}_2)\right]
\Bigr\}\,,
\nn
&&f_{\rm II}^{(p)}= {1\over 2} e^{-2u_0 }\dot{\phi}^2\Bigl\{ \Bigr.\mu \Bigl[
 \bigr.
(p-1)_2 A_p +q_1 A_q +2\left((p-1)(p+2)\dot{u}_1^2+(3p-1)q\dot{u}_1\dot{u}_2
+q(q+1)\dot{u}_2^2\right)
\nn
&&
\hs{20}+ 4\dot{\phi}\left((p-1)\dot{u}_1+q\dot{u}_2\right)\bigl.\Bigr]
+2\dot{\phi}\left[(2\nu+\lambda)\dot{\phi}
+ 2\nu\left(p\dot{u}_1+q\dot{u}_2\right)\right]\Bigl.\Bigr\}\,,
\nn
&&f_{\rm II}^{(q)}= {1\over 2}
 e^{-2u_0 }\dot{\phi}^2\Bigl\{ \Bigr.\mu
\Bigl[ \bigr. p_1
A_p +(q-1)_2 A_q +2\left(p(p+1)\dot{u}_1^2+p(3q-1)\dot{u}_1\dot{u}_2+(q-1)(q+2)
\dot{u}_2^2\right)
\nn
&&\hs{20}+ 4 \dot{\phi}\left(p\dot{u}_1+(q-1)\dot{u}_2\right)\bigl.\Bigr]
+2\dot{\phi}\left[(2\nu+\lambda)\dot{\phi}
+ 2 \nu\left(p\dot{u}_1+q\dot{u}_2\right)\right]\Bigl.\Bigr\}\,,
\nn
&&f_{\rm II}^{(\phi)}=\frac{1}{8}
e^{-2u_0}\dot{\phi}\left\{\mu\left[ p_1A_p(\dot{\phi}+2\dot{u}_1)+
q_1A_q(\dot{\phi}+2\dot{u}_2)+2pq\dot{u}_1\dot{u}_2(\dot{u}_1+\dot{u}_2
+\dot{\phi})\right]\right.\nn
&&\hs{20}\left.+2\nu\dot{\phi}\left[p(p+1)\dot{u}_1^2+2pq\dot{u}_1\dot{u}_2
+q(q+1)\dot{u}_2^2\right]
+8(\lambda+\nu)(p\dot{u}_1+ q \dot{u}_2)\dot{\phi}^2+2(3\lambda+2\nu)
\dot{\phi}^3
\right\}\,,\nn
&&g_{\rm II}^{(p)}=-\mu  (p-1)   e^{-2u_0} \dot{\phi}^2\,,\quad
 g_{\rm II}^{(q)}
=-\mu  (q-1)
 e^{-2u_0 } \dot{\phi}^2\,,\quad
g_{\rm II}^{(\phi)}=-\frac{p}{4} e^{-2u_0} \dot{\phi}\left[\nu \dot{\phi}+\mu \left(
(p-1)\dot{u}_1+q\dot{u}_2\right)\right]\,,
\nn
&&h_{\rm II}^{(p)}=-\mu  q  e^{-2u_0} \dot{\phi}^2\,,\quad
h_{\rm II}^{(q)}=-\mu  p
 e^{-2u_0 } \dot{\phi}^2\,,\quad
h_{\rm II}^{(\phi)}=-\frac{q}{4} e^{-2u_0} \dot{\phi}\left[\nu \dot{\phi}
+\mu \left( p\dot{u}_1+(q-1)\dot{u}_2\right)\right]\,,
\nn
&&i_{\rm II}^{(p)}= 2  e^{-2u_0} \dot{\phi}\left[\nu \dot{\phi}
+\mu ((p-1)\dot{u}_1+q\dot{u}_2)\right]\,,\quad
i_{\rm II}^{(q)}= 2 e^{-2u_0} \dot{\phi}\left[\nu \dot{\phi}
+\mu (p\dot{u}_1+(q-1)\dot{u}_2)\right]\,,\nn
&&
i_{\rm II}^{(\phi)}=-\frac{1}{8}e^{-2u_0}\left\{\mu \Bigl[p_1A_p+ q_1A_q
+2 pq
\dot{u}_1\dot{u}_2 \Bigr]+4\dot{\phi}\left[(2\nu +3\lambda)\dot{\phi}+\nu\left( p\dot{u}_1
+q\dot{u}_2\right)\right]\right\}\,,
\label{gb3}
\ena
and
\begin{eqnarray*}
R^2_{\rm GB}&=& e^{-4u_0} \Big\{ p_3 A_p^2 + 2p_1 q_1 A_p A_q + q_3 A_q^2
+ 4 \dot u_1 \dot u_2 (p_2 q A_p + p q_2 A_q) + 4 p_1 q_1 \dot u_1^2 \dot u_2^2 \nn
&+&4pX \left[ (p-1)_2 A_p + q_1 A_q + 2(p-1)q \dot u_1 \dot u_2 \right]
~~~
 +\; 4qY \left[ p_1 A_p + (q-1)_2 A_q
 + 2p(q-1) \dot u_1 \dot u_2 \right]\Big\}.
\end{eqnarray*}

We have four basic equations~\p{rfe1}--\p{rfe4} for three variables $u_1, u_2$
and $\phi$. However those four equations are not independent.
In fact, Eq.~\p{rfe1} is a constraint equation, which has no second order
 derivative,
and four functionals satisfy the following identity:
\bea
\dot {\cal F} +(p \dot u_1 + q \dot u_2 -2 \dot u_0-2 \dot\phi) {\cal F}
= p \dot u_1 {\cal F}^{(p)} + q \dot u_2 {\cal F}^{(q)}
+8\, \dot \phi {\cal F}^{(\phi)}\,.
\label{Bianchi}
\ena
As a result, three of them become independent, which we have to solve.
In order to discuss the dynamics of the present system, we first
analyze the fixed points, which turn out to be important.

\vskip 1cm

\end{widetext}
\section{accelerating universes and their stability}
\subsection{Fixed Point Solutions}
\label{FP_solution}

Here we study the properties of the fixed point solutions
 in the basic equations \p{rfe1}--\p{rfe4}.
In this paper, we restrict our analysis to the simple case of flat
$p$- and $q$- spaces ($\sigma_p=\sigma_q=0$).
We choose $u_0=0$ by use of the gauge freedom of time coordinate.
In this case, Eqs.~\p{rfe1}--\p{rfe4} form
an autonomous system for three variables:
\bea
{\it \Theta}=\dot u_1\,,~~
\theta=\dot u_2\,, ~~
{\rm and}~~~
\varpi=\dot \phi
\,.
\ena

The fixed point is given by setting the three variables to be constants
such that ${\it \Theta}={\it \Theta}_0$, $\theta=\theta_0$,
and $\varpi=\varpi_0$. Hence the equations for the fixed points are given by
\begin{eqnarray*}
F({\it \Theta}_0, \theta_0,\varpi_0)&\equiv&
{\cal F}|_{{\it \Theta}={\it \Theta}_0,
\theta=\theta_0,\varpi=\varpi_0}=0, \nn
F^{(p)}({\it \Theta}_0, \theta_0,\varpi_0)&\equiv&
{\cal F}^{(p)}|_{{\it \Theta}={\it \Theta}_0,
\theta=\theta_0,\varpi=\varpi_0}=0, \nn
F^{(q)}({\it \Theta}_0, \theta_0,\varpi_0)&\equiv&
{\cal F}^{(q)}|_{{\it \Theta}={\it \Theta}_0,
\theta=\theta_0,\varpi=\varpi_0}=0, \nn
F^{(\phi)}({\it \Theta}_0, \theta_0,\varpi_0)&\equiv&
{\cal F}^{(\phi)}|_{{\it \Theta}={\it \Theta}_0,
\theta=\theta_0,\varpi=\varpi_0}=0
\,.
\nonumber
\end{eqnarray*}

Before giving the fixed points explicitly, let us discuss the properties of the solutions.
Since $\dot u_1={\it \Theta}_0$, $\dot u_2=\theta_0$, and
$\dot \phi=\varpi_0$, we find the metric and dilaton field as
\begin{eqnarray*}
u_1&=&{\it \Theta}_0 \, t+{\rm constant}\,,\nn
u_2&=& \,\theta_0 \, t\, +{\rm constant} \,, \nn
\phi&=&\varpi_0 \, t+ {\rm constant}\,.
\nonumber
\end{eqnarray*}
The integration constants can be absorbed in rescaling of
the $p$-spatial coordinates $\vect{x}$
and $q$-spatial coordinates $\vect{y}$ as well as in
a translational shift  of time coordinate $t$.
As a result, we find the metric  and the dilaton field as
\begin{eqnarray*}
ds_{D}^2&=&-dt^2+e^{2{\it \Theta}_0 t}d\,\vect x^2+
e^{2\theta_0 t}d\,\vect y^2
\,,\nn~~
\phi&=&\varpi_0 t \,.
\end{eqnarray*}

The gravity of our universe, which is obtained by compactification
of $D$-dimensional
spacetime, must be described by the four-dimensional Einstein-Hilbert action,
which means that the Newtonian gravitational ``constant'' is really constant.
Hence we have to extract the Einstein frame for $(p+1)$-dimensional spacetime.
Using this frame description, we can discuss the accelerating expansion
of the Universe
or inflation.
The  $(p+1)$-dimensional Einstein frame metric
$ds_E^2$ is extracted from the total $D$-dimensional spacetime as
\begin{eqnarray*}
ds_D^2= \exp{\left[-\frac{2(qu_2-2\phi)}{p-1}\right]}ds_E^2+e^{2u_2}ds_q^2
\end{eqnarray*}
with
\begin{eqnarray*}
~~~~~~
ds_E^2=\exp{\left[\frac{2(qu_2-2\phi)}{p-1}\right]}\left(-dt^2+e^{2u_1}ds_p^2
\right).
\end{eqnarray*}
The contribution from the dilaton $\phi$ is due to the non-minimal coupling
between the dilaton and gravity in $D$-dimensional spacetime
(see Eq.~(\ref{gali})).
Introducing the cosmic time $\tau$ in  $(p+1)$-dimensional spacetime by
\begin{eqnarray*}
\tau&=&\int \exp{\left[\frac{(q\theta_0-2\varpi_0)}{p-1}\, t \right]}dt
\nn
&=&\left\{ \begin{array}{cc}
\left(\frac{p-1}{q\theta_0-2\varpi_0}\right)
\exp{\left[\frac{(q\theta_0-2\varpi_0)}{p-1}\, t \right]}
 & (q\theta_0\neq2\varpi_0) \\[1em]
t & (q\theta_0= 2\varpi_0) \\
\end{array} \right.
\,,
\nonumber
\end{eqnarray*}
we find the Friedmann-Lema\^itre-Robertson-Walker
 (FLRW) spacetime;
\bea
ds_E^2=-d\tau^2+a^2(\tau)ds_p^2
\,,
\ena
where
\begin{eqnarray*}
a=&&
\left\{ \begin{array}{cccc}
 \left[\left({q\theta_0-2\varpi_0\over p-1}\right)\tau\right]^{P}
 &
 & (q\theta_0\neq 2\varpi_0)\\[1em]
e^{{\it \Theta}_0 \, \tau }
 &&  (q\theta_0= 2\varpi_0) \\
\end{array} \right.
\,.
\\[.5em]
&&{\rm with}
~~~
P=1+{(p-1){\it \Theta}_0\over q\theta_0-2\varpi_0}\,.
\end{eqnarray*}

Note that if $q\theta_0> 2\varpi_0$, then $\tau=0\rightarrow \infty$ as
$t=-\infty\rightarrow \infty$,
while  if $q\theta_0< 2\varpi_0$, then $\tau=-\infty\rightarrow 0$
 as
$t=-\infty\rightarrow \infty$.
For the case of $q\theta_0=2\varpi_0$,
we find $\tau=t=-\infty \rightarrow \infty$.

In the case of $q\theta_0\neq 2\varpi_0$,
the condition for an expanding universe is given by
\bea
H\equiv {da/ d\tau \over a}= {1\over \tau}{d\ln a\over  d\ln \tau}
={P\over \tau}>0
\,,
\ena
while the condition for an accelerating universe is given by
\bea
\frac{1}{a}{d^2a\over d\tau^2}&=&{1\over \tau^2}\left[
{d^2\ln a\over  (d \ln \tau)^2} +
\left({d\ln a\over  d\ln \tau}\right)^2-{d\ln a\over  d\ln \tau}\right]
\nn
&=&{P(P-1)\over \tau^2}>0
\,,
\ena
which are rewritten by the fixed point variables as
\bea
&&[(p-1){\it \Theta}_0+q\theta_0-2\varpi_0]>0\,,
\label{acond1}
\\
&&
{\it \Theta}_0[(p-1){\it \Theta}_0+q\theta_0-2\varpi_0]>0
\,,
\label{acond2}
\ena
respectively.
As a result, the accelerating expansion of the universe in $(p+1)$-dimensional
spacetime is obtained for the case of $\Theta_0>0$ and
$[(p-1){\it \Theta}_0+q\theta_0-2\varpi_0]>0$.

\vskip 1.3cm
\subsection{The accelerating expansion of the Universe in 4-dimensional
spacetime}

Now we solve the equation for fixed points.
In what follows, we restrict our analysis to
the ten-dimensional string theory with our 3-space, i.e.,
 $p=3$, $q=6$, and $D=10$.
We also normalize the time coordinate $t$ (or $\tau$) by
$\alpha_2^{-1/2}=2\sqrt{2}(\alpha')^{-1/2}$.

The ten-dimensional metric is given by
\begin{eqnarray*}
ds_{10}=&& b^{-{2(3\theta_0-\varpi_0)\over \theta_0}}ds_E^2+b^2ds_6^2
\,,
\\[.5em]
{\rm with}&&
ds_E^2=-d\tau^2+a^2 ds_3^2
\,.
\end{eqnarray*}
In the case of $3\theta_0\neq \varpi_0$, the metric components
are given by the cosmic time $\tau$ as

\bea
a~\propto~\tau^P\,,~~{\rm and}~~~
b~\propto~\tau^Q\,,
\ena
where
\bea
P={1+{{\it \Theta}_0\over 3\theta_0-\varpi_0}}
~~{\rm and}~~~Q={\theta_0\over 3\theta_0-\varpi_0}
\,.
\ena

The conditions for the accelerating expansion of the Universe are
\bea
({\it \Theta}_0+3\theta_0-\varpi_0)>0
\,~~{\rm and}~~~
{\it \Theta}_0>0
\,.
\label{Accelexp}
\ena
Note that $\tau=0\rightarrow \infty$ for $ 3\theta_0>\varpi_0$, while
$\tau=-\infty\rightarrow 0$ for $ 3\theta_0<\varpi_0$.
For the case of $3\theta_0=\varpi_0$, we find an exponential de Sitter expansion,
 $a\propto \exp[{\it \Theta}_0 \tau]$ with ${\it \Theta}_0>0$.
The cosmic time in the four-dimensional Einstein frame is the same as that
in the ten-dimensional string frame.

\begin{widetext}

The equations for fixed points to be solved are
\bea
F&=&6 \left({\it \Theta}_0+\theta_0\right)\left({\it \Theta}_0+5\theta_0\right)
-12 ({\it \Theta}_0+2 \theta_0) \varpi_0
+4 \varpi_0^2
\nn
&+&24 \left[2{\it \Theta}_0^3 (3 \theta_0- \varpi_0)
+9 {\it \Theta}_0^2 \theta_0
 (5 \theta_0-4 \varpi_0)
+30 {\it \Theta}_0 \theta_0^2 (2 \theta_0-3
\varpi_0)+5 \theta_0^3 (3 \theta_0-8 \varpi_0)\right]
\nn
&-&9 ({\it \Theta}_0+\theta_0) ({\it \Theta}_0+5 \theta_0) \varpi_0^2 \mu-2 \varpi_0^3
(3 {\it \Theta}_0+6 \theta_0+\varpi_0)\nu
-3 \varpi_0^4
\lambda =0\,,
\label{F_eq}
\\
F^{(p)}&=&2 \left[3 ({\it \Theta}_0^2+4 {\it \Theta}_0 \theta_0+7 \theta_0^2)
-4 ({\it \Theta}_0+3 \theta_0) \varpi_0
+2 \varpi_0^2\right]
\nn
&+&8 \left[12 {\it \Theta}_0^3 \theta_0+81 {\it \Theta}_0^2 \theta_0^2
+180 {\it \Theta}_0 \theta_0^3+105 \theta_0^4 \right.
\nn
&&\qquad \qquad \qquad \left.-4 \left({\it \Theta}_0^3+15 {\it \Theta}_0^2 \theta_0
+51 {\it \Theta}_0 \theta_0^2+45 \theta_0^3\right)
 \varpi_0 +4 \left({\it \Theta}_0^2+12 {\it \Theta}_0 \theta_0+15 \theta_0^2\right)
 \varpi_0^2\right]
\nn
&+&\varpi_0^2 \left[-3 \left({\it \Theta}_0^2+4 {\it \Theta}_0 \theta_0+7 \theta_0^2\right)
+4 ({\it \Theta}_0+3 \theta_0) \varpi_0\right]
\mu +2 \varpi_0^4 \nu +\varpi_0^4 \lambda =0\,,
\label{Fp_eq}
\\
F^{(q)}&=&2 \left(6 {\it \Theta}_0^2+15 {\it \Theta}_0 \theta_0+15 \theta_0^2
-6 {\it \Theta}_0 \varpi_0
-10 \theta_0 \varpi_0+2 \varpi_0^2
\right)
\nn
&+&8 \left\{3 {\it \Theta}_0^4+9 {\it \Theta}_0^3 (5 \theta_0-2 \varpi_0)
+30 {\it \Theta}_0\theta_0
\left(5 \theta_0^2-7 \theta_0 \varpi_0+2
\varpi_0^2\right)\right.
\nn
&&\qquad\qquad\qquad \left.+3 {\it \Theta}_0^2 \left(45 \theta_0^2
-40 \theta_0 \varpi_0+4 \varpi_0^2
\right)
+5 \theta_0^2 \left(9 \theta_0^2-20 \theta_0 \varpi_0+8 \varpi_0^2\right)
\right\}
\nn
&+&\varpi_0^2 \left\{-6 {\it \Theta}_0^2+5 \theta_0 (-3 \theta_0+2 \varpi_0)
+{\it \Theta}_0 (-15 \theta_0+6 \varpi_0)\right\} \mu
+2 \varpi_0^4 \nu+\varpi_0^4 \lambda =0\,,
\label{Fq_eq}
\\
F^{(\phi)}&=&2 \left(6 {\it \Theta}_0^2+18 {\it \Theta}_0 \theta_0
+21 \theta_0^2-6 {\it \Theta}_0 \varpi_0-12 \theta_0
\varpi_0+2 \varpi_0^2
\right)
\nn
&+&24 \left({\it \Theta}_0^4+18 {\it \Theta}_0^3 \theta_0+66 {\it \Theta}_0^2 \theta_0^2
+90 {\it \Theta}_0 \theta_0^3+35\theta_0^4\right)
\nn
&+&3 \varpi_0 ({\it \Theta}_0+\theta_0) ({\it \Theta}_0+5 \theta_0)
(3 {\it \Theta}_0+6 \theta_0-\varpi_0)\mu
\nn
&&\qquad\qquad\qquad
+(9 ({\it \Theta}_0+2\theta_0)^2
\varpi_0^2-2
\varpi_0^4)\nu-3 \varpi_0^3 (-2 {\it \Theta}_0-4\theta_0+\varpi_0)\lambda=0\,,
\label{Fphi_eq}
\ena
where the first, second and third rows in the equations
are obtained from the lowest Lagrangian ${\cal L}_{\bf 0}$,
the GB term ${\cal L}_{\rm I}$,
and the other correction term ${\cal L}_{\rm II}$, respectively.

Since these equations are not independent, but
are related with each other by one constraint \p{Bianchi},
we need three of them to be solved.
Although all these three equations are complicated, we find one simple equation
from $F^{(p)}=0$ and $F^{(q)}=0$, i.e.,
\bea
&&F^{(q)}-F^{(p)}\nn
&&=\left({\it \Theta}_0-\theta_0\right)
\left(3 {\it \Theta}_0+6 \theta_0-2 \varpi_0\right)
\left(2+8 {\it \Theta}_0^2
+80 {\it \Theta}_0 \theta_0+80 \theta_0^2
-32 {\it \Theta}_0 \varpi_0-80 \theta_0 \varpi_0-\varpi_0^2 \mu\right)
\nn
&&=0.
\label{Fqp}
\ena
Hence, we adopt $F=0$, $F^{(p)}=0$ and Eq. (\ref{Fqp})
as three independent equations~\cite{foonote2}.
{}From Eq.~(\ref{Fqp}), we classify the fixed points into three cases:
\bea
&&{\it 1}.~~{\it \Theta}_0=\theta_0,
\label{F1_eq}
\\[.5em]
&&{\it 2}.~~3 {\it \Theta}_0+6 \theta_0-2 \varpi_0=0,
\label{F2_eq}
\\[.5em]
&&{\it 3}.~~2+8 {\it \Theta}_0^2
+80 {\it \Theta}_0 \theta_0+80 \theta_0^2
-32 {\it \Theta}_0 \varpi_0-80 \theta_0 \varpi_0-\varpi_0^2 \mu=0
\label{F3_eq}
\,.
\ena

To solve these equations, we introduce new variables $r=\theta_0/
{\it \Theta}_0$ and $s=\varpi_0/{\it \Theta}_0$, assuming ${\it \Theta}_0
\neq 0$.
For the case of ${\it \Theta}_0=0$, see  Appendix \ref{append1}.
We then obtain from $F=0$ and $F^{(p)}=0$
\bea
&&2 {\it \Theta}_0^{-2}\left[
3(1+r)(1+5r)-6(1+2r)s+2s^2\right]
\nn
&&~~~+24 \left[3r(2+15r+20r^2+5r^3)-2s(1+18r+45r^2+20r^3)+2s^4\right]
\nn
&&~~~-3s^2[3(1+r)(1+5r)-2s^2]\mu-2 s^3
(3+6r-2s)\nu
=0\,,
\label{F_rs}
\\
&&2{\it \Theta}_0^{-2}\left[
3(1+4r+7r^2)-4(1+3r)s+2s^2\right]
\nn
&&~~~+8 \left[3r(4+27r+60r^2+35r^3)
-4(1+15r+51r^2+45r^3)s+4(1+12r+15r^2)s^2-2s^4\right]
\nn
&&~~~-s^2 \left[3 \left(1+4 r+7 r^2\right)
-4 (1+3 r) s +2 s^2\right]
\mu  =0\,,
\label{Fp_rs}
\ena
where we have used $\lambda=-2(\mu+\nu)-16$.
Eqs.~(\ref{F1_eq}), (\ref{F2_eq}) and (\ref{F3_eq}) are rewritten as
\bea
&&{\it 1}.~~r=1,
\label{F1_rs}
\\[.5em]
&&{\it 2}.~~3 +6 r-2 s=0,
\label{F2_rs}
\\[.5em]
&&{\it 3}.~~2{\it \Theta}_0^{-2} +8
+80 r+80 r^2
-32 r-80 rs-s^2 \mu=0
\label{F3_rs}
\,.
\ena

Note that there always exists a trivial fixed point, i.e.,
${\it \Theta}_0=\theta_0=\varpi_0=0$, which corresponds to
a Minkowski spacetime with a constant dilaton.
Next we shall find non-trivial fixed points in the above three cases separately.

\subsubsection{${\it \Theta}_0=\theta_0 ~(r=1)$}

First we investigate the case of ${\it \Theta}_0=\theta_0~(r=1)$,
which means both our 3-space and internal 6-space expand with the same
expansion rate. It gives a ten-dimensional de-Sitter solution in
the string frame. However, the dilaton field is not always trivial.
As a result, we find that our 4-dimensional spacetime is either de Sitter universe
for the case of $3\theta_0= \varpi_0~(s=3)$,
or  the power-law expanding universe with the power
$P=(4-s)/(3-s)$ for $3\theta_0\neq \varpi_0~(s\neq 3)$.

It follows from Eqs.~(\ref{F_rs}) and (\ref{Fp_rs})
 that we have to solve
\bea
&&2{\it \Theta}_0^{-2}\left(
18-9s+s^2\right)
+24 \left(63-84s+s^4\right)
-3s^2(18-s^2)\mu- s^3
(9-2s)\nu
=0\,,
\label{F_rs1}
\\
&&2{\it \Theta}_0^{-2}\left(
18-8s+s^2\right)
+8 \left(189
-224s+56s^2-s^4\right)
-s^2 \left(18-8 s+s^2\right)
\mu  =0\,,
\label{Fp_rs1}
\ena
which are solved as
\bea
&&
2 (16 + 2 \mu + \nu) s^5 - (264 + 41 \mu + 25 \nu) s^4 +
  4 (32 + 27 \mu + 27 \nu) s^3
\nn
&&
~~~~
+ 2 (1904 + 63 \mu - 81 \nu) s^2 -
  72 (112 + 9 \mu) s - 2520 =  0,
\label{eq_s1}
\\
&&
{\it \Theta}_0^2=-
{2 \left(
18-8s+s^2\right)\over 8 \left(189
-224s+56s^2-s^4\right)
-s^2 \left(18-8 s+s^2\right)
\mu}
\,.
\label{eq_Th1}
\ena
Solving Eq.~(\ref{eq_s1}) for $s$ for given coupling constants $\mu$ and $\nu$
and inserting the real valued solution into
Eq.~(\ref{eq_Th1}), we find the values of fixed points, i.e.,
$({\it \Theta}_0, \theta_0, \varpi_0)=(1,1,s){\it \Theta}_0$.
$\mu$ and $s$ must satisfy the following inequality:
\bea
\mu>{8 \left(189
-224s+56s^2-s^4\right)\over s^2 \left(18-8 s+s^2\right)}
\,,
\label{4.36}
\ena
to find the real valued solution for ${\it \Theta}_0$
since $(18-8s+s^2)>0$ for real $s$.
Note that there is always positive ${\it \Theta}_0$ as a solution
for \p{eq_Th1} when this condition is satisfied.

Now the conditions for the accelerating expansion are given in \p{Accelexp},
which translate into ${\it \Theta}_0>0$ and $s<4$ in this case.
There are several fixed points satisfying these criteria.
Checking the behavior of the right hand side of \p{4.36}, we find that
it has a minimum $\sim -18.0961$ at $s=\frac12 (7-\sqrt{4\sqrt{7}-7})\sim 2.55$
and grows to infinity near the origin, and then monotonically decreases to $-8$
for negative large $s$.
It follows that in order to get accelerating expansion, $\mu$ must be larger
than the minimum value  $\sim -18.0961$. For $-8>\mu \geq -18.0961$,
the allowed $s$ has values in the range $4>s>0$.
For $\mu>-8$, we can also have a solution $s<0$.
Given $\mu$ in this range, $s$ should be chosen such that it satisfies \p{4.36}
and then ${\it \Theta}_0$
and $\nu$ can be determined by \p{eq_Th1} and \p{eq_s1},
respectively.

In order to find de Sitter solution in four dimensions,
we have to set $s=3$, for which Eqs.~(\ref{eq_s1}) and (\ref{eq_Th1}) yield
\bea
3\mu+\nu+32=0\,,~~{\rm and}~~~
{\it \Theta}_0=\pm \sqrt{
{2\over 9\mu+160}}
\,.
\label{eq_pi}
\ena
which requires $\mu>-160/9$.
The fixed point corresponding to de Sitter expanding universe is
\bea
({\it \Theta}_0, \theta_0, \varpi_0)=
\left(
\sqrt{
{2\over 9\mu+160}},\sqrt{
{2\over 9\mu+160}},3\sqrt{
{2\over 9\mu+160}}
\right)\,.
\ena
Note that such a solution is not found without taking account of field redefinition
ambiguity because the necessary condition is $3\mu+\nu+32=0$.
Although we find de Sitter universe in four dimensions,
it cannot be our universe because the internal space is also expanding exponentially.
The result is summarized in Table~\ref{t1}.

\begin{table}[h]
\begin{center}
\begin{tabular}{|c@{}c@{}|@{}c@{}|c|c|} \hline \hline
~~case & &
fixed point $({\it \Theta}_0,\theta_0,\varpi_0)$~~~~~~~~~~
 & $H={\it \Theta}_0$~~~~~~ &
$\nu$~~~~~~~~~~~~~~~~~~~~~~~~~~
\\ \hline \hline
&&&& \\[-.4em]
{\it 1.}&$[{\it \Theta}_0=\theta_0]$~~~~~~~~~~~~~~~~~~~
& $({\it \Theta}_0,{\it \Theta}_0,3{\it \Theta}_0)$
~~~~~~~~~~~~~~~~~~~~~~
& $\displaystyle{\pm \sqrt{2\over 9\mu+160}}$~~ &
$-(3\mu+32)$~~~~~~~~~~~~~~   \\[1em] \hline
&&&& \\[-.6em]
{\it 2}.&$[3 {\it \Theta}_0+6 \theta_0-2 \varpi_0=0]$~~
&---~~~~~~~~~~~~~~~~~~~~~~~~~~&---&---~~~~~~~~~~~~~~~~~
\\[.2em] \hline
&&&& \\[-.4em]
{\it 3}.&$[2(1+4 {\it \Theta}_0^2
-8 {\it \Theta}_0 \theta_0$~~~~~~&
~~$({\it \Theta}_0,-2.94771{\it \Theta}_0,-8.84313{\it \Theta}_0)$~~
&~~$\displaystyle{\pm {0.159922 \over \sqrt{\mu+17.0724}}}$~~~
& $- 3.86891 \mu-45.4052$~~
 \\[1em]
\cline{3-5}
&&&& \\[-.4em]
&$-80 \theta_0^2)=9\theta_0^2 \mu)]$
&$({\it \Theta}_0,0.583777{\it \Theta}_0,1.75133{\it \Theta}_0)$~~~ &
$\displaystyle{\pm {0.807509  \over \sqrt{\mu+ 18.2148}}}$
& $- 3.40790 \mu -39.2874$~~
 \\[1em] \hline
\end{tabular}
\caption{The fixed points of de Sitter spacetime $a=e^{Ht}$ in four dimensions.
There exists no de Sitter solution in the case ${\it 2}$. In the cases of
{\it 1} and {\it 3}, we find one-parameter family for de Sitter solutions
($\mu$ is a free parameter). The other coupling constant $\nu$ is fixed by
$\mu$ as given in the last column.
The scale factor of the internal space  and the evolution of the dilaton are
 given by $b=e^{\theta_0 t}$ and $\phi=\varpi_0 t$, respectively.
The expanding universe in four dimensions ($H>0$) is stable against
small perturbations. The realistic
inflationary solution is given by ${\it \Theta}_0>0$ and $\theta_0<0$,
which is found in the case {\it 3}.}
\label{t1}
\end{center}
\end{table}

We show other fixed points for $(\mu,\nu)=(-15.4,14.1), (-12,4)$ and
$(0, 48.2)$, which are solved numerically, and their properties are given
in Table \ref{t2} as examples.

\begin{table}[h]
\begin{center}
\begin{tabular}{|c||c|c|@{}c@{}|@{}r@{}|c|@{}r@{}|c|} \hline \hline
case &$\mu$&$\nu$& ~~fixed point (${\it \Theta}_0,\theta_0,\varpi_0$) &
 ~~$P$~~~~~~~~&~~A/D~~& ~~$M_0$~~~~~~ &~stability~\\ \hline \hline
{\it 1}.&$-15.4$&$14.1$
&$(0.307622,0.307622,0.903627)$& $16.9893$~~
& A  &$-0.961344$~~&S\\ \cline{4-8}
&&&$(-0.250813, -0.250813, -1.30743)$& $0.548081$~~
&D & ~$-0.357553$~~&S\\  \cline{2-8}
&$-12$&$4 $&$(-0.118465,-0.118465,-0.685402)$& $0.641022$~~
&D & $ -0.304618$~~&S\\  \cline{2-8}
&$0$&$48.2 $&$(-0.0787943, -0.0787943, -0.346152)$& $ 0.016844$~~&D
& $0.282184$~~&US \\  \hline
{\it 2}.&$-$&$-$&
$-$& $-$~~~~~~~& $-$
&$-$~~~~~~~~&$-$\\ \hline
{\it 3}.&$-15.4$&$14.1$&
$(0.107856, -0.364542, -1.10847)$& $8.26718$~~
&A & $-0.353253$~~&S\\ \cline{4-8}
&&&$(0.0060359, 0.431902, 0.727932)$& $1.01063$~~
&A & $-1.15366$~~&S\\ \cline{4-8}
&&&$(0.948509, -0.0160689, 0.334445)$& $-1.47878$~~
&A & $-2.08022$~~&S\\ \cline{4-8}
&&&$(-0.765888, 0.0881743, -1.01054)$& $0.399332$~~
&D &$-0.252459$~~&S\\
\cline{2-8}
&$-12$&$4$&$(~0.909500,-0.063855,~0.198882)$& ~$-1.329387$~~
&A & $-1.947607$~~&S\\ \cline{4-8}
&$ $&$ $&$(~0.323252,-0.235434,-0.517073)$& $-0.708264$~~
&A & $-0.591299$~~&S\\ \cline{4-8}
&$ $&$ $&$(-0.567964,~0.235739,-0.347272)$& $0.461385$~~
&D &$-0.405089$~~&S\\ \cline{4-8}
&$ $&$ $&$(-0.035670,~0.325903,~0.508717)$& $ 0.923944$~~
&D & $- 0.830977$~~&S\\    \cline{2-8}
&$0$&$48.2 $&$(0.7982, -0.166337, -0.107164)$& $-1.03702$~~
&A & $ -1.61091$~~&S\\  \cline{4-8}
&$ $&$ $&$(~0.895263, -0.117903, 0.0561875)$& $-1.18412$~~
&A & $-1.86599$~~&S\\  \cline{4-8}
&$ $&$ $&$(-0.101297, -0.0676883, -0.346229)$& $0.292438$~~
&D & $0.0175641$~~&US\\  \cline{4-8}
&$ $&$ $&$(-0.111046, -0.0629703, -0.346321)$& $0.294546$~~
&D & $0.0183167$~~&US\\  \cline{4-8}
&$ $&$ $&$(-0.344151, 0.332526, 0.169237)$& $0.968383$~~
&D & $-0.624232$~~&S\\  \cline{4-8}
&$ $&$ $&$(-0.505956, 0.318319, -0.0787433)$& $1.05549$~~
&D & $-0.549534$~~&S\\ \hline
\end{tabular}
\caption{The examples of the fixed points for some values of
$\mu$ and $\nu$. For each pair of $\mu$ and $\nu$, we find two types of
fixed points; one corresponds to the  power-law expanding spacetime
in four dimensions ($a=\tau^P$) and the other is the contracting one, which is
just a time reversal solution of the expanding spacetime.
We show the expanding universes. A and D denote the accelerating or decelerating
universe, respectively. Some of them show the accelerating expansion with
large positive power-exponent, while some others show the similar accelerating
behaviour to the solution found in the Einstein-Gauss-Bonnet-dilaton system
(see Appendix \ref{append2} and \cite{BGO}).
S and US denote ``stable" and ``unstable" against small perturbations,
respectively. The decelerating spacetimes with the positive power exponent
are unstable. The others including accelerating spacetimes are stable.}
\label{t2}
\end{center}
\end{table}

\subsubsection{$3 {\it \Theta}_0+6 \theta_0-2 \varpi_0=0~(3+6r-2s=0)$}

In this case,
there exists only a trivial fixed point
$({\it \Theta}_0,\theta_0,\varpi_0)=(0,0,0)$
for any values of $\mu$ and $\nu$, as follows:
Inserting ${\it \Theta}_0=-2(3 \theta_0- \varpi_0)/3$
into Eqs.~(\ref{F_eq}) and (\ref{Fp_eq}), we find
\bea
&&
6 (27 \theta_0^2 - 12 \theta_0 \varpi_0 + 2 \varpi_0^2) -
 8 (729 \theta_0^4 - 540 \theta_0^3 \varpi_0 + 648 \theta_0^2 \varpi_0^2
 - 240 \theta_0 \varpi_0^3 + 38 \varpi_0^4)
\nn
&&~~~~
 - 9 \mu \varpi_0^2 (27 \theta_0^2 - 12 \theta_0 \varpi_0 + 2 \varpi_0^2)=0,
\nn
&&
- 18 (27 \theta_0^2 - 12 \theta_0 \varpi_0 + 2 \varpi_0^2) +
 8 (729 \theta_0^4 - 540 \theta_0^3 \varpi_0 + 648 \theta_0^2 \varpi_0^2
- 240 \theta_0 \varpi_0^3 + 38 \varpi_0^4)
\nn
&&~~~~
+ 9 \mu \varpi_0^2 (27 \theta_0^2 - 12 \theta_0 \varpi_0 + 2 \varpi_0^2)=0
\,,
\ena
which imply
\bea
27 \theta_0^2 - 12 \theta_0 \varpi_0 + 2 \varpi_0^2=0
\,.
\ena
This equation has no real root unless $\theta_0= \varpi_0=0$.
As a result, we have only the trivial fixed point
$({\it \Theta}_0,\theta_0,\varpi_0)=(0,0,0)$.

\subsubsection{$2+8 {\it \Theta}_0^2
+80 {\it \Theta}_0 \theta_0+80 \theta_0^2
-32 {\it \Theta}_0 \varpi_0-80 \theta_0 \varpi_0-\varpi_0^2 \mu=0$\\[.5em]
{\rm (}$2{\it \Theta}_0^{-2}-s^2 \mu+8
+80 r+80 r^2
-32 s-80 rs$=0 {\rm )}}

In this last case, there are also several fixed points for
various values of $\mu$ and $\nu$.
We find an
interesting fixed point, which describes de-Sitter or rapidly accelerating
expanding universe for a finite range of $(\mu,\nu)$ parameter space.
It also shows a nice property such as dynamical compactification
of higher-dimensional spacetime, i.e.,
our 3-space is expanding while the internal space shrinks.

The equations for fixed points (\ref{F_rs}) and (\ref{Fp_rs})
are rewritten as
\bea
&&\left(2{\it \Theta}_0^{-2}-s^2\mu\right)\left[
3(1+r)(1+5r)-6(1+2r)s+2s^2\right]
\nn
&&~~~+24 \left[3r(2+15r+20r^2+5r^3)-2s(1+18r+45r^2+20r^3)+2s^4\right]
\nn
&&~~~-2s^2\mu[3(1+r)(1+5r)+3(1+2r)s-4s^2]-2 s^3 \nu
(3+6r-2s)
=0\,,
\label{F_rs_3}
\\
&&\left(2{\it \Theta}_0^{-2}-s^2\mu\right)\left[
3(1+4r+7r^2)-4(1+3r)s+2s^2\right]
\nn
&&~~~+8  [3r(4+27r+60r^2+35r^3)
-4(1+15r+51r^2+45r^3)s\nn
&&~~~+4(1+12r+15r^2)s^2-2s^4 ]
 =0\,.~~~~~~
\label{Fp_rs_3}
\ena
Inserting the present condition,
\bea
2{\it \Theta}_0^{-2}-s^2 \mu=-8(1 +10 r+10 r^2 -4 s-10 rs)\,,
\ena
into those two equations, we find the equation for $r$ and $s$:
\bea
&&
2 s^4 - 4 (2 + 5 r) s^3 + 2 (7 + 30 r + 40 r^2) s^2 -
 2 (2 + 5 r) (3 + 10 r + 15 r^2) s
\nn
&&~~~+
 3 (1 + 10 r + 30 r^2 + 50 r^3 + 35 r^4)
=0,
\label{F_deSitter}
\\
&&
2 (16 + 2  \mu +\nu)  s^3 -
  3 (\mu + \nu) (1 + 2 r)
   s^2 - [3 (16 +  \mu) + 2 (136 + 9  \mu) r + 15 (16 +  \mu) r^2] s \nn
&&~~~
- 8 r (2 + 5 r) =0.
\label{Fp_deSitter}
\ena

First we  try to find de Sitter solution, which is given by
the condition of $3\theta_0=\varpi_0$.
Eqs.~(\ref{F_deSitter}), (\ref{Fp_deSitter}) and (\ref{F3_eq})
are reduced to
\bea
&&
8 (6 s^2- 107 s -60) + 3 (s^2- 27 s -9  ) \mu - 27 s\nu=0,
\label{eq_mn}
\\
&&
s^4+ 6 s^3- 18 s^2+ 54 s-81=0,
\label{eq_s}
\\
&&
{\it \Theta}_0^2={18\over 8(20 s^2+6 s-9)+9s^2 \mu}
\label{eq_Theta}
\,,
\ena
where
$s\equiv \varpi_0/{\it \Theta}_0$.
Solving Eq.~(\ref{eq_s}),
we find two real solutions
\bea
s=\left\{
\begin{array}{l}
s_1
\equiv {3\over 2} \left(-1 -\eta - \sqrt{7 -  \eta^2 + 10 /\eta}\right)=
-8.84313
\\
s_2
\equiv {3\over 2} \left(-1 -\eta + \sqrt{7 -  \eta^2 + 10 /\eta}\right)=
1.75133
\\
\end{array}
\right.
\,,
\ena
where
\bea
\eta= {1\over \sqrt{3}}\left[7 - 2 \left({11 + 3 \sqrt{69}\over 2}\right)^{1/3} +
     2 \left({-11 + 3 \sqrt{69}\over 2}\right)^{1/3}\right]^{1\over 2}
=1.36393527 .
\ena

The fixed point is given by
$({\it \Theta}_0,\theta_0,\varpi_0)=(1,s/3,s){\it \Theta}_0$,
where ${\it \Theta}_0$ is given by Eq.~(\ref{eq_Theta}).
To find the real solution, we have to require that
\bea
\mu>-{8(20 s^2+6 s-9)\over 9s^2}=\left\{
\begin{array}{rl}
-17.0724 & (s=s_1)  \\
-18.2148 & (s=s_2)  \\
\end{array}
\right.
\,.
\ena
We then find
\bea
{\it \Theta}_0=
\left\{
\begin{array}{l}
\pm {0.159922 \over \sqrt{\mu+17.0724 }}
\\[.5em]
\pm {0.807509  \over \sqrt{\mu+ 18.2148}}
\\
\end{array}
\right.
\,,~~~
\nu=\left\{
\begin{array}{l}
- 3.86891 \mu-45.4052
\\[.5em]
- 3.40790 \mu -39.2874
\\
\end{array}
\right.
\,,~~~
\begin{array}{l}
{\rm for}~~~s=s_1
\label{eq_mn1}
\\[.5em]
{\rm for}~~~s=s_2
\,.
\end{array}
\ena

The positive ${\it \Theta}_0$
gives de Sitter expansion of our 3-space.
However, the internal space is also expanding in the  case of $s>0$,
which is not realistic. Hence the fixed point corresponding to
the de Sitter expanding universe
in four dimensions with the contracting internal space is given by
\bea
({\it \Theta}_0,\theta_0,\varpi_0)
={1\over \sqrt{\mu+17.0724}}
\left(0.159922 ,
-0.471405,
-1.41421 \right)
\label{FT_dS}
\,,
\ena
where $\mu(>-17.0724)$ is a free parameter.
This solution gives
\bea
a=\exp[{\it \Theta}_0t]\,,~~b=\exp[\theta_0t]\,,~~{\rm and}
~~~~e^\phi=\exp[\varpi_0t]\,,
\ena
with the values (\ref{FT_dS}) at the fixed point.
The Hubble parameter of the de Sitter solution is
\bea
H\equiv {\it \Theta}_0={0.159922 \over \sqrt{\mu+17.0724}}
~~\alpha_2^{-{1\over 2}}={0.452328 \over \sqrt{\mu+17.0724}}
~~(\alpha')^{-{1\over 2}}
\,.
\ena
The results for the de Sitter solutions are summarized in Table~\ref{t1}.

For the other fixed points, we solve the equations numerically.
For some values of $\mu$ and $\nu$,
we present numerical values of the fixed points and their properties
in Table \ref{t2}.
We show the contour of the power exponent $P$ of the
four-dimensional scale factor $a=\tau^P$ in Fig.~\ref{fig1}.
In the $(\mu,\nu)$ parameter plane,
we depict the contours of $P=1$ and $10$ as well as
the lines (\ref{eq_mn1}) for de Sitter solution.
We find that there exists a finite range, which gives the accelerating
expansion of the universe.

\begin{figure}[H]
\centering
\includegraphics[width=6cm,height=5cm,clip]{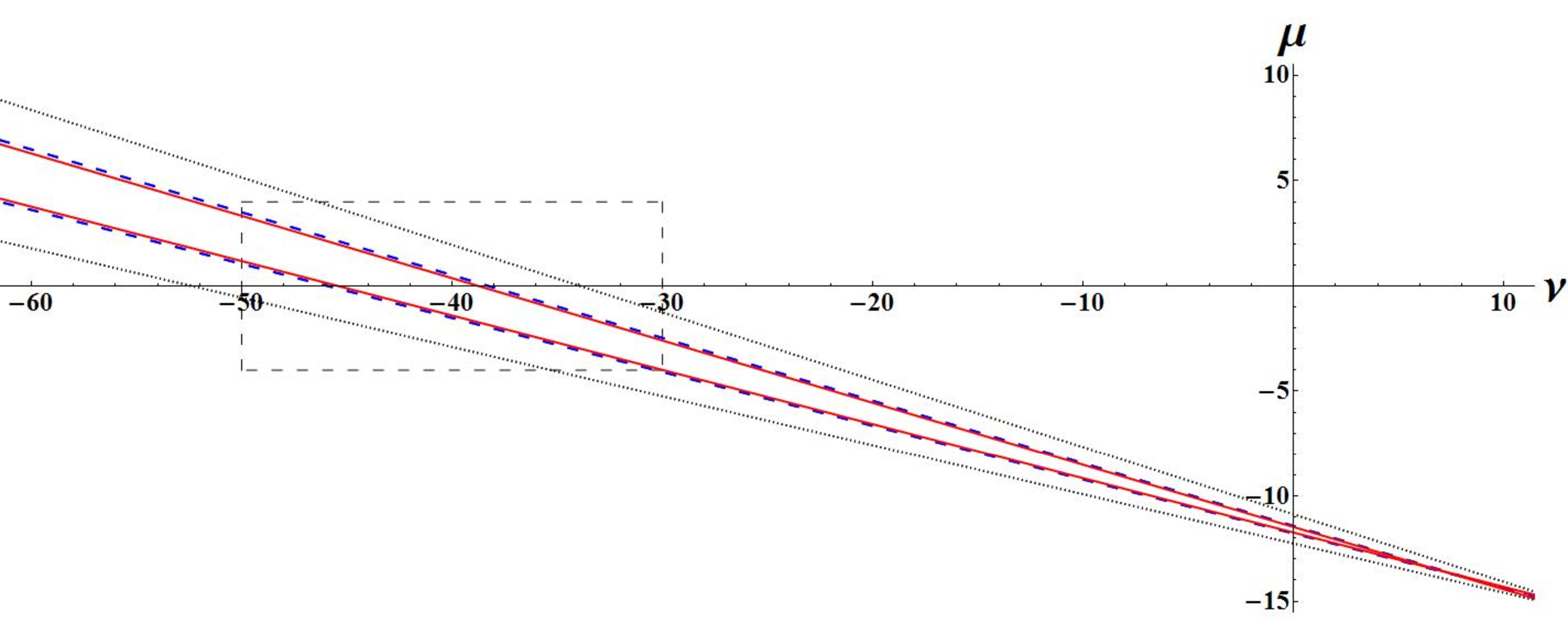}
\hskip 2cm
\includegraphics[width=5cm,height=5cm,clip]{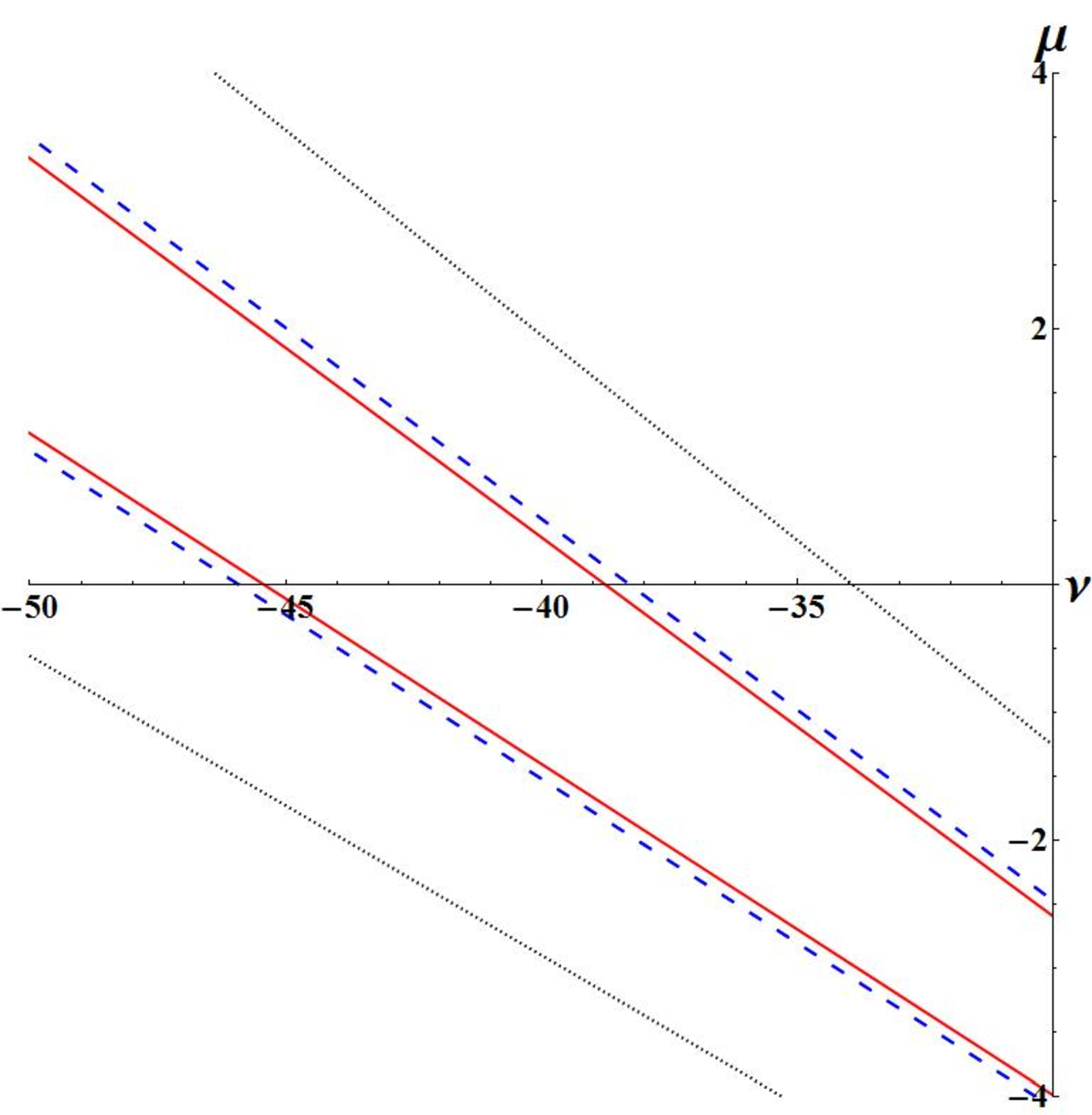}
\caption{The contours of the power exponent $P$ of the
4-dimensional scale factor $a(t)\sim \tau^P$.
The (black) dotted, (blue) dashed and (red) solid  curves
correspond to  the contours of $P=1$,
$P=10$, and $P=\infty$ (de Sitter solution), respectively.
The right is the magnification of the the dotted square region
in the left figure.
The accelerated expanding universe with $P>1$ exist in the region
between two contour curves of $P=1$ and $P=\infty$.
The solutions with
the large negative power exponents ($P\ll -1$)
lie near de Sitter solution in the region
between two (red) solid lines.
 }
\label{fig1}
\end{figure}

\end{widetext}

\subsection{Stability Analysis}

The analysis of stability for the fixed points is important because
it predicts which fixed point is naturally
realized in the present dynamical system.
Hence we investigate stability of the fixed points found above.

We perturb Eqs.~(\ref{rfe1})--(\ref{rfe4})
 around the fixed point as
${\it \Theta}={\it \Theta}_0+ \delta {\it \Theta}$,
$\theta =\theta_0 + \delta \theta$, and
$\varpi=\varpi_0+\delta \varpi$.
The equations for the perturbed variables, $\delta {\it \Theta},
\delta {\theta}$ and $\delta {\varpi}$, are obtained  as
\bea
{d\over dt}
\left(
\begin{array}{c}
\delta {\it \Theta} \\
\delta \theta\\
\delta \varpi \\
\end{array}
\right)
={\cal M}_0\,
\left(
\begin{array}{c}
\delta {\it \Theta} \\
\delta \theta\\
\delta \varpi \\
\end{array}
\right)
\,,
\ena
where ${\cal M}_0$ is a $3\times 3$ matrix, which is evaluated at
the fixed point $({\it \Theta}_0, \theta_0, \varpi_0 )$.
Although there are four basic equations (\ref{rfe1})--(\ref{rfe4}),
one of them is derived from the other three.
As a result, we have three independent basic equations.
It is why we find a $3\times 3$ matrix in the perturbation equations.

Although the perturbation equations are very complicated,
those evaluated at the fixed points turn out to be remarkably simple:
\bea
{\cal M}_0
=M_0 \,
\left(
\begin{array}{ccc}
1&0&0\\
0&1&0\\
0&0&1 \\
\end{array}
\right)
\,,
\ena
where
we find only one degenerate eigenvalue
$M_0=-(3{\it \Theta}_0+6\theta_0-2\varpi_0)$~\cite{footnote}.
If $M_0>0$, the fixed point is unstable, while if
$M_0<0$, it is stable. $M_0=0$ gives a marginal stability.
As we discussed in Sec.~\ref{FP_solution},
the condition for accelerating expansion of the three-dimensional space
is given by ${\it \Theta}_0>0$ and $({\it \Theta}_0+3\theta_0-\varpi_0)>0$.
As a result, if the expanding universe is
accelerated, then
the present fixed points are stable ($M_0<0$).
If the expansion is decelerated
($0<P<1$), we find two cases;
the  stable and unstable ones.
The expansion is decelerating  if
\bea
0<P={{\it \Theta}_0+3\theta_0-\varpi_0\over 3\theta_0-\varpi_0}<1
\, ,
\ena
which yields $(3\theta_0-\varpi_0)>0$ and ${\it \Theta}_0<0$
for expanding solutions satisfying (4.5).
Hence the stability condition is rewritten as
$(3+6r-2s)<0$, i.e., $(3r-s)<-3/2$, which corresponds to
 the power exponent $1/3<P (<1)$.
On the other hand, if
\bea
-{3\over 2}<3r-s<-1 ~~{\rm or}~~~0<P<1/3,
\ena
is satisfied, the decelerated expanding universe
becomes unstable (See the example with $\mu=0$ and $\nu=48.2$
in Table \ref{t2}.)

This stability condition $(3{\it \Theta}_0+6\theta_0-2\varpi_0)>0$
is understandable because it means
that the ten-dimensional
``volume" in the string frame ($\sqrt{-g}e^{-2\phi}$) is expanding.

From the four-dimensional point of view,
we may understand the above instability as follows:
If the Universe contains a stiff matter fluid, which
equation of state is given by ${\cal P}=\rho$,
the power exponent of the scale factor is given by $P=1/3$.
Here ${\cal P}$ and $\rho$ are
the pressure of fluid and its energy density.
The causality condition, under which
the sound velocity is less than the velocity of light,
 implies ${\cal P}\leq \rho$, which corresponds to
the equation of state with
$P\geq 1/3$.
If the power exponent of the universe
is given by $P<1/3$, we have the effective
fluid with the  equation of state ${\cal P}>\rho$,
 which violates the causality condition.
It may be the reason why we find instability for $P<1/3$.

\section{Conclusion and Discussion}

In this paper, we have studied cosmological solutions in the effective theory with
$\alpha'$-correction terms in the string frame.
Since there exists an ambiguity in the effective theory
in the first order of the corrections via field redefinition,
we analyze generalized effective theories obtained by field redefinition.
We restrict our analysis to the effective theories, which yield equations of motion
without derivatives of order higher than two just as ``Galileon" field theory.
We find that the effective action of such theories contains two additional correction terms
$ \mu \left(R^{AB}-{1\over 2}Rg^{AB}\right)
\nabla_A\phi\nabla_B\phi$ and $\nu \Box\phi(\nabla\phi)^2$
as well as the GB combination $R_{\rm GB}^2$
and $\lambda (\nabla\phi)^4$, where
$\mu$ and $\nu$ are free coupling constants while
$\lambda$ is fixed as $\lambda=-2(\mu+\nu)-16$.

We find de Sitter solution as well as the power-law expanding universe in
four-dimensional spacetime.
The accelerated expanding spacetime is always stable.
The de Sitter solution is given by one-parameter family of solutions because
$\mu$ and $\nu$ must satisfy some relation.
The Hubble expansion scale is $H\sim 0.1 (\alpha')^{-1/2}$ if the coupling
constant $\mu\sim O(1)$. Since $(\alpha')^{-1/2}\sim M_{10}$, where
$M_{10}$ is ten-dimensional Planck mass, $H$ can be much smaller than
the 4-dimensional Planck mass, if the extra dimension is large enough.
We also find the accelerating expansion of the universe with
a positive power-exponent for a finite range of the ($\mu,\nu$)-space.
Those solutions as well as de Sitter universe in four dimensions
have not been found in the effective theories without field redefinition.
As shown in Appendix \ref{append2}, we do not find so large difference between
cosmological behaviour in the effective theory in the string frame and that
in the Einstein frame. Hence we conclude that
it is very important to take into account
field redefinition when we discuss macroscopic objects such as the universe.

Our present effective action may be too simple to discuss a realistic cosmology.
However it is important for us to find de Sitter expanding or accelerating
universe in four-dimensional Einstein frame.
Since such spacetimes are attractors in the present dynamical system and then
stable, a rapidly accelerating universes are naturally realized if
the coupling constants satisfy appropriate conditions.
Although de Sitter or nearly de Sitter universe will be naturally found,
there is no way out from such a rapid expansion because of stability.
Toward a realistic inflationary scenario, we have to find how to finish
this rapid expansion.
We can suppose that there may be the following two ways:
One is via changes of $\mu$ and $\nu$ as the running coupling constants
of the renormalization group.
When the universe evolves, a typical energy scale will change.
As a result, the coupling constants will also change.
If $\mu$ and $\nu$ move from the rapid expansion range,
then inflation will end.
The other possibility is stabilization
of the dilaton field and moduli field.
In the present model, we have not considered any mechanism to stabilize
those fields. In fact, they are time-dependent in most cosmological solutions.
However, if they are changing in time now, the fundamental constants may
become time-dependent, which is inconsistent with many
experiments and observation.
Hence those scalar fields must be  fixed at some stage of cosmological
evolution by unknown mechanism.
It might be related to supersymmetry breaking via gaugino condensation.
Although we do not know the precise mechanism,
it will change the dynamics of the dilaton $\phi$ and the moduli $b$.
As a result, our dynamical system will be completely changed.
Once those fields are fixed, de Sitter or nearly de Sitter expansion
is no longer possible. The inflationary phase must cease.

Even if we find the way out from inflation, we still have several problems
to establish our inflationary scenario.
We have to find a reheating mechanism and the origin of density fluctuation.
We may need second inflation for such purposes.

\section*{Acknowledgments}

We would like to thank N. Sakai, Y. Tanii and A. Tseytlin for discussions.
Part of this work was carried out while the authors were
attending Summer Institute 2010 (Cosmology \& String).
We thank the organizers for their hospitality and support
by Grant-in-Aid for Creative Scientific Research No. 19GS0219.
This work was also supported in part by the Grant-in-Aid for
Scientific Research Fund of the JSPS (C) No. 20540283,
No. 21$\cdot$09225, No. 22540291 and (A) No. 22244030.

\begin{widetext}

\appendix
\section{Fixed points in some special cases}
\label{append1}

In this Appendix, we present the fixed points explicitly
for some special cases in which one of ${\it \Theta}_0, \theta_0,
$ and $\varpi_0$ vanishes.

\subsection{${\it \Theta}_0=0$}
\label{sub1}
\noindent
{\it 1}. (${\it \Theta}_0=\theta_0=0$)
\\
\hskip 0.5cm
We find
${\it \Theta}_0=\theta_0=0, \varpi_0= \pm \sqrt{2/(\mu+8 )}$
only if $\mu>-8$ and $2 \mu + \nu + 16 = 0$.\\
{\it 2}. ($3\theta_0-
\varpi_0=0$)\\
\hskip 0.5cm
There is no solution. \\
{\it 3}. ($2+80 \theta_0^2
-80 \theta_0 \varpi_0-\varpi_0^2 \mu=0$)\\
\hskip 0.5cm
We find
\bea
\theta_0 &&
= \pm \frac{ \sqrt{2\left(95+18 \sqrt{15}-5 \sqrt{85+78 \sqrt{15}}\right)}}
{\sqrt{5\left[147 \mu +8 \left(335-\sqrt{15}-\sqrt{-8300+2270 \sqrt{15}}\right)\right]}}
=\pm {0.424874\over \sqrt{\mu+16.81379}}
~~~~~~~
\nn
\varpi_0
&&
= \pm \frac{7 \sqrt{6}}{\sqrt{147 \mu+8 \left(335-\sqrt{15}-\sqrt{-8300+2270 \sqrt{15}}\right)}}
=\pm
{1.41421\over \sqrt{\mu+16.81379}}
\ena
with
\bea
\nu&=&-\frac{1}{14} \left[16 \left(30+ 2\sqrt{15}+\sqrt{-20+22 \sqrt{15}}\right)
+\left(44+ 2\sqrt{15}+\sqrt{-20+22 \sqrt{15}}\right) \mu \right]
\nn
&=&
-(4.27293 \mu +52.3668)
\label{mn1}
\ena
and
\bea
\theta_0 &&
= \pm \frac{\sqrt{2 \left(95+18 \sqrt{15}+5 \sqrt{85+78 \sqrt{15}}\right)}}
{\sqrt{5\left[147 \mu+8 \left(335-\sqrt{15}+\sqrt{-8300+2270 \sqrt{15}}\right)\right]}}
=\pm {0.846099\over \sqrt{\mu+19.227249}}
~~~~~~~
\nn
\varpi_0
&&
= \pm \frac{7 \sqrt{6}}{\sqrt{147 \mu+8 \left(335-\sqrt{15}+\sqrt{-8300+2270 \sqrt{15}}\right)}}
=\pm
{1.41421\over \sqrt{\mu+19.227249}}
\ena
with
\bea
\nu&=&-\frac{1}{14} \left[16 \left( 30+2\sqrt{15}-\sqrt{-20+22 \sqrt{15}}\right)
+\left(44+2\sqrt{15}-\sqrt{-20+22 \sqrt{15}}\right) \mu\right]
\nn
&=&
-(3.11935 \mu + 33.9097)
\label{mn2}
\ena

These solutions exist if $\mu$ and $\nu$ satisfy the condition
(\ref{mn1}) or (\ref{mn2}), which means
they form one-parameter family.
Since ${\it \Theta}_0=0$, the scale factor in four-dimensional spacetime is
given by $a=\tau$, which corresponds to the Milne universe.
The dynamics of the internal space or the dilaton field
induces the dynamics in four dimensions.
\subsection{${\theta}_0=0$}
\label{sub2}
\noindent
{\it 1}. (${\it \Theta}_0=\theta_0=0$)
\\
\hskip 0.5cm
It is the same as the case {\it 1} in Sec.~\ref{sub1}.
 \\
{\it 2}. ($3{\it \Theta}_0-
2\varpi_0=0$)\\
\hskip 0.5cm
There is no solution. \\
{\it 3}. ($2+8 {\it \Theta}_0^2
-32 {\it \Theta}_0 \varpi_0-\varpi_0^2 \mu=0$)\\
\hskip 0.5cm
We can reduce the basic equations as follows:
\bea
&&
2s^4 - 8 s^3 + 14 s^2 - 12  s + 3   = 0,
\label{eq_xi}
\\
&&4 (6s^4 + 8  s^3 - 26 s^2 + 12  s -3  )
+ (4s^2- 3  s -3 )s^2 \mu  + (2s -
     3 )s^3 \nu = 0,
\label{eq_xi_mn}
\\
&&{\it \Theta}_0=\pm  \sqrt{2\over (\mu s^2 + 32 s -8 ) }
\,,
\label{eq_xi_Th}
\ena
where $s=\varpi_0/{\it \Theta}_0$.
Solving Eq.~(\ref{eq_xi}) for $\xi$, we find two real solutions;
\bea
s=\left\{
\begin{array}{l}
s_1\equiv {1\over 2}\left(2 - \sqrt{2 (-1 + \sqrt{3})}\right)=0.395000\\
s_2\equiv {1\over 2}\left(2 + \sqrt{2 (-1 + \sqrt{3})}\right)=1.605000\\
\end{array}
\right. .
\ena

For $s=s_1$,
\bea
&&
\nu= -\left[\left(4 + \sqrt{3}-\sqrt{1 + \sqrt{3}}\right) \mu
 +16 \left(3 + \sqrt{3}
- \sqrt{1 + \sqrt{3}}\right)\right]
= -(4.07916 \mu+49.2665),
\nn
&&
{\it \Theta}_0=\pm
{\sqrt{2}\left(\sqrt{3} + 1
+\sqrt{ (1 + \sqrt{3})}\right) \over \sqrt{3 \mu +
 8 \left[7 + \sqrt{3} - 2 \sqrt{1 + \sqrt{3}}
 + 2 \sqrt{3 (1 + \sqrt{3})}\right]}}
=\pm{3.58029 \over \sqrt{\mu+29.73881}}
\,,\nn
&&
\varpi_0=\pm {\sqrt{6}\over \sqrt{3 \mu +
 8 \left[7 + \sqrt{3} - 2 \sqrt{1 + \sqrt{3}}
 + 2 \sqrt{3 (1 + \sqrt{3})}\right]}}
=\pm{1.41421 \over\sqrt{\mu+29.73881}},
\ena
and
for $s=s_2$, we obtain
\bea
&&
\nu=-\left[\left(4 + \sqrt{3}+\sqrt{1 + \sqrt{3}}\right) \mu
 +16 \left(3 + \sqrt{3}
+ \sqrt{1 + \sqrt{3}}\right)\right]=-(7.38494 \mu +102.15908),
\nn
&&
{\it \Theta}_0=\pm
{\sqrt{2}\left(\sqrt{3} + 1
-\sqrt{ (1 + \sqrt{3})}\right) \over \sqrt{3 \mu +
 8 \left[7 + \sqrt{3} + 2 \sqrt{1 + \sqrt{3}}
 - 2 \sqrt{3 (1 + \sqrt{3})}\right]}}
=\pm{0.88113 \over \sqrt{\mu+16.8321}}
\,,\nn
&&
\varpi_0=\pm {\sqrt{6}\over \sqrt{3 \mu +
 8 \left[7 + \sqrt{3} + 2 \sqrt{1 + \sqrt{3}}
 - 2 \sqrt{3 (1 + \sqrt{3})}\right]}}
=\pm{1.41421 \over\sqrt{\mu+16.8321}} .
\ena

This fixed point is also one-parameter family.
The power exponent $P$ of the universe is
\bea
P=
\left\{
\begin{array}{l}
-{1\over \sqrt{3}} - \sqrt{{1\over 3} (1 + \sqrt{3})}=-1.53165
\\[.5em]
-{1\over \sqrt{3}} + \sqrt{{1\over 3} (1 + \sqrt{3})}=~~0.376947
\\
\end{array}
\right. .
\ena
for $s=s_1$ and $s_2$, respectively.
The former case gives accelerating universe as $\tau\rightarrow 0$.
However the string coupling constant $g=e^\phi$ diverges
because $\varpi_0>0$ if the 3-space is
expanding (${\it \Theta}_0>0$).
The unknown stringy effect should be taken into account
in the limit of $\tau\rightarrow 0$.

\subsection{$\varpi_0=0$}
\label{AppA3}
There exist no solutions for the cases {\it 1} and {\it 2}.
For the case {\it 3}, we find
\bea
&&
35 r^4 + 50 r^3 + 30 r^2+ 10 r +1  =0,
\label{eq_r}
\\
&&
{\it \Theta}_0 =\pm  {1\over 2\sqrt{-( 10 r^2+ 10 r +1 )}}
\,,
\label{eq_Th}
\ena
where $r=\theta_0/{\it \Theta}_0$.
Solving Eq.~(\ref{eq_r}), we find two real roots;
\bea
r=\left\{
\begin{array}{l}
r_1=-{1\over 14} \left[5 + \xi
+ \sqrt{{(2 - \xi) (13 + 2 \xi + \xi^2)\over \xi}}\right]
=-0.706924
\\
r_2=-{1\over 14} \left[5 + \xi
 - \sqrt{{(2 -\xi) (13 + 2 \xi+ \xi^2)\over \xi}}\right]
=-0.156263
\end{array}
\right.
\,,
\ena
where
\bea
\xi = \left({2\over 5}\right)^{1/3} \sqrt{7
\left[ (5 + 3 \sqrt{5})^{1/3} - (-5 + 3 \sqrt{5})^{1/3}\right]
 - 3}=1.0423107
\,.
\ena
Those roots give the fixed points:
\bea
{\it \Theta}_0=
\left\{
\begin{array}{l}
\pm 0.482957
\\
\pm 0.886034
\\
\end{array}
\right.
\,,~~{\rm and}~~~\theta_0=
\left\{
\begin{array}{l}
\mp 0.341414
\\
\mp 0.138454
\\
\end{array}
\right.
\,,
\label{sol_varpi0}
\ena
for $r=r_1$ and $r_2$, respectively.

However it turns out that this is not the solution in the present system,
because it does not satisfy the last equation (\ref{Fphi_eq}).
Although (\ref{sol_varpi0}) satisfies three equations;
$F=0$, $F^{(p)}=0$ and $F^{(q)}=0$, they do not guarantee
$F^{(\phi)}=0$ unless $\dot \phi\neq 0$ (see Eq.~(\ref{Bianchi})).
As a result, (\ref{sol_varpi0}) is no longer the fixed point in our system.
We find only a trivial fixed point of $(0,0,0)$.

Note that this solution (\ref{sol_varpi0})
was  found by Ishihara for the ten-dimensional
Einstein-Gauss-Bonnet model without a dilaton field~\cite{ISHI}.
It gives us some caution that
by assuming that a dilaton field is constant,
the solutions in the theory with a dilaton field
are not always given by
the solutions in the theory
without a dilaton field.

\vskip .5cm

\section{Cosmological solutions for the Einstein-Gauss-Bonnet-dilaton system
in  string frame}
\label{append2}
\end{widetext}

In order to discuss the frame dependence, we analyze cosmological models
in the Einstein-Gauss-Bonnet-dilaton system in the string frame.
The low-energy effective action with the GB correction term
in a general frame is given by
\bea
S&=&\frac{1}{2\kappa_D^2}\int d^Dx \sqrt{-\tilde g} \left\{
e^{\a \phi} \left[ R + \beta (\pa_A \phi)^2 \right]
\right.
\nn
&&
\left.
~~~~~~~~~~~~~~~~~~~~~~~~
+ \a_2 e^{-\gamma \phi} R^2_{\rm GB}\right\}
\,,~~~~~
\label{gf}
\ena
where we drop the higher-order corrections of the dilaton field $\phi$.
The choice of $\a=-2,\b=4, \c=2$ corresponds to the action in
the string frame, whereas $\a=0,\b=-1/2, \c=\sqrt{2/(D-2)}$ gives
that in the Einstein frame.
Although the descriptions in two frames are related via a conformal
transformation~\cite{conformal_transformation},
the effective theories in both frames are different,
because we do not include the higher-order corrections of the dilaton field~\cite{conf}.
So, in this appendix,  we will analyze cosmological models in the effective
theory in the string frame to
see the difference between two effective theories.
The effective theory in the Einstein frame was analyzed in~\cite{BGO}.
Note that if  we do not include a dilaton field,
which was analyzed by Ishihara~\cite{ISHI}, both frame descriptions
are equivalent and then the solutions must be reduced to those found in~\cite{ISHI}.

Let us consider the metric in $D$-dimensional space,
which metric is assumed to be Eq.~(\ref{metric}).
We then find the basic equations given by Eqs.~(\ref{rfe1})--(\ref{rfe4})
with $\mu=\nu=\lambda=0$. Note that this effective theory is not
involved in our family of effective theories discussed in the text,
because  $\mu=\nu=\lambda=0$ does not satisfy
the constraint between coupling constants, $\lambda+2(\mu+\nu)+16=0$.
Here, we also consider only flat internal and external spaces
($\sigma_p=\sigma_q=0$) for simplicity.
We also set $u_0=0$ by using the gauge freedom.
Then the basic equations (\ref{rfe1}) -(\ref{rfe4})
turn out to be  an autonomous system for the variables
${\it \Theta} \equiv \dot{u}_1$,
 $\theta \equiv \dot{u}_2$, $\varpi \equiv \dot{\phi}$, where
${\it \Theta}$ and $\theta$ denote
the expansion parameter for $p$-space  and $q$-space, respectively.

\begin{widetext}

\begin{figure}[H]
\centering
\includegraphics[width=7cm,height=5.5cm,clip]{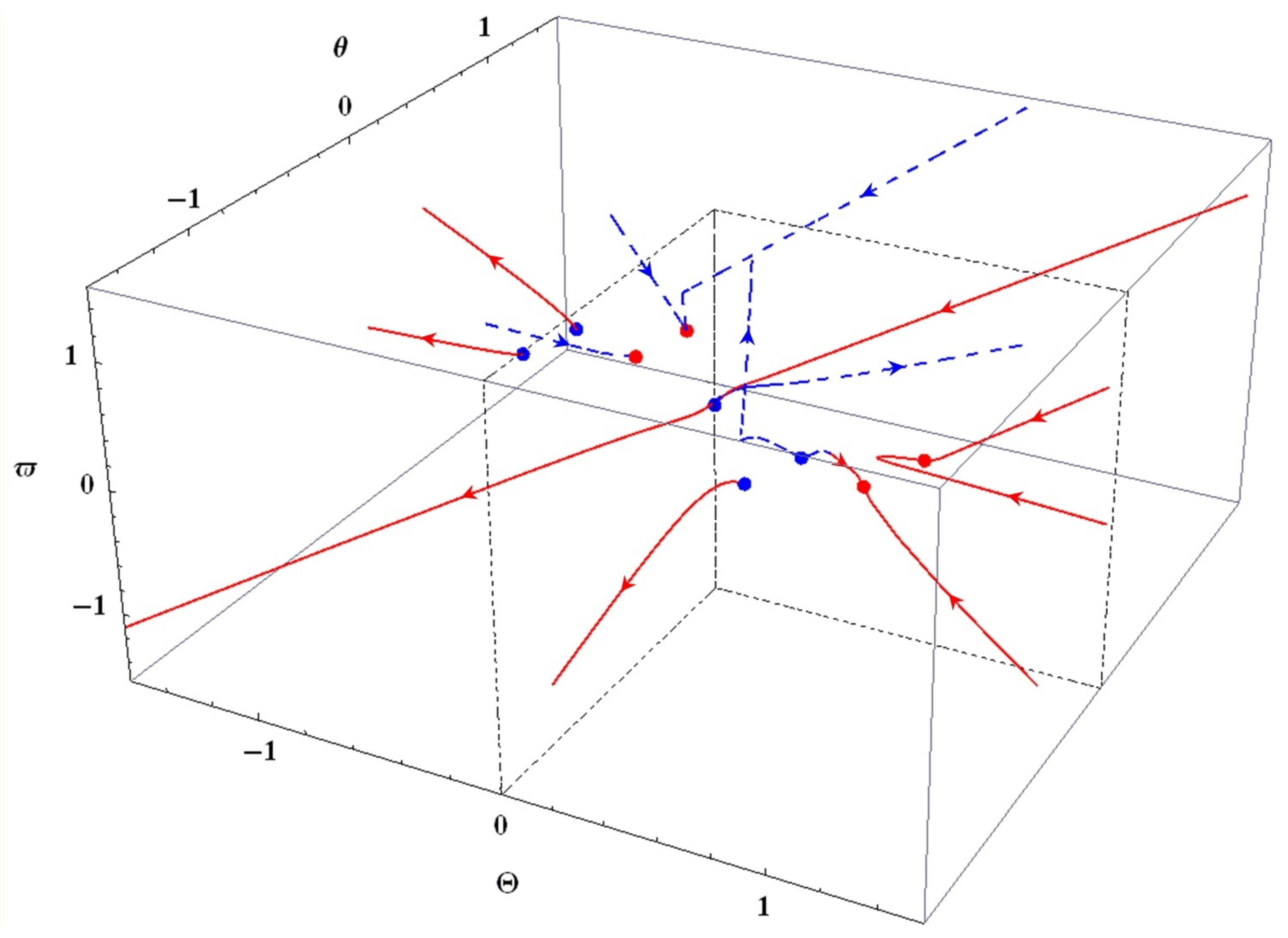}
\hskip 1cm
 \includegraphics[width=7cm,height=5.5cm,clip]{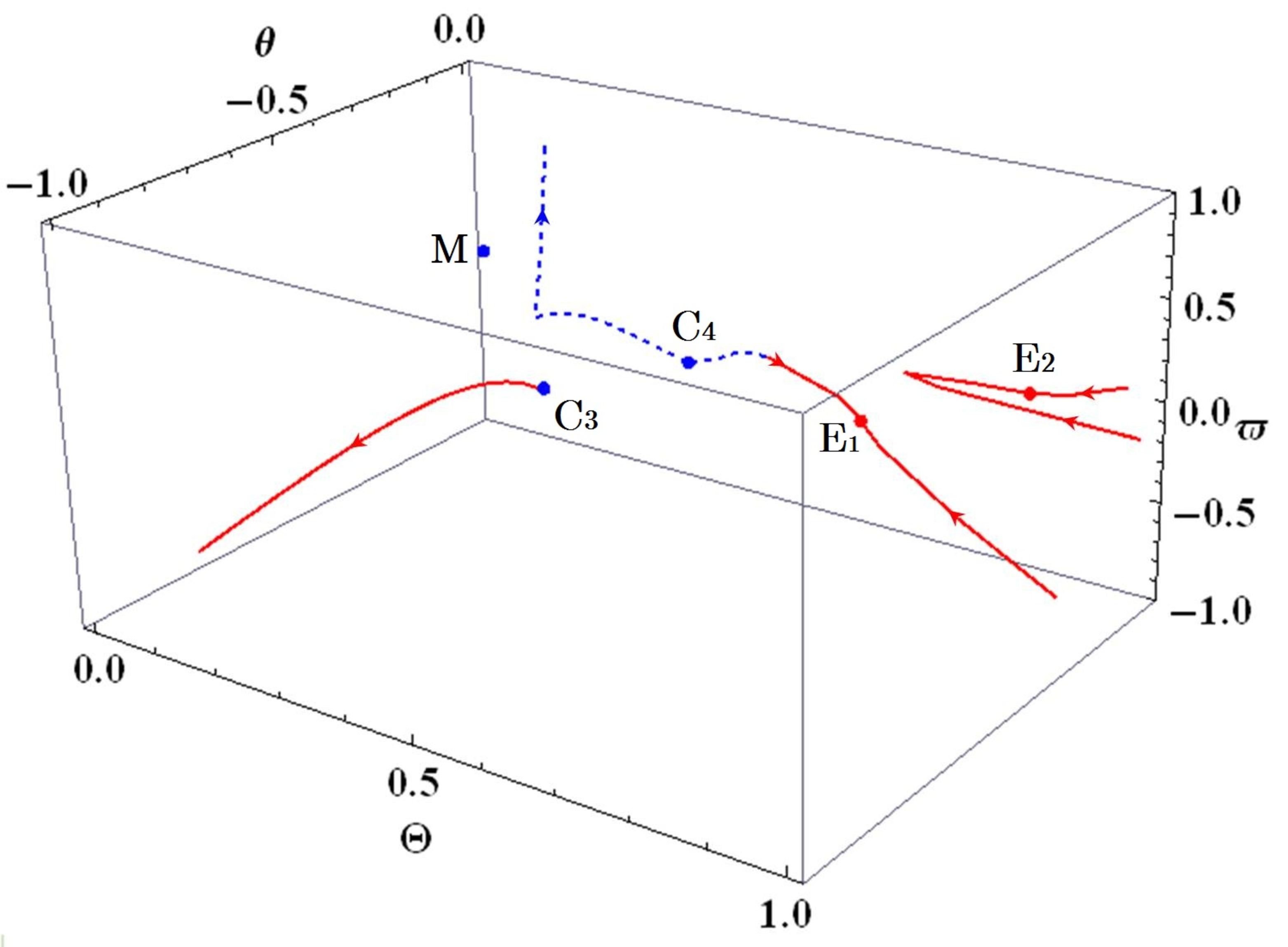}
\caption{The orbits of cosmological solutions, which
evolutionary directions are shown by the arrows.
In the right, we also show the magnification  of the most interesting quarter
(the dotted region in the left figure),
in which our 3-space is expanding
while the internal space is contacting.
The red (blue) points correspond to the stable expanding
universe (unstable contracting universe).
The fixed points of the expanding universes are attractors, while
those of the contracting universes are repellers.
 The solid (red) and dashed (blue) curves
denote the orbits of the
accelerating and decelerating periods of the universe,
respectively.
There exists one interesting orbit, which starts from
unstable C$_4$ and ends at stable E$_1$.
}
\label{dynamics_EGBD}
\end{figure}
\end{widetext}

Eq.~\p{rfe1} is a constraint equation, in which there is no
derivative of those three variables.
Any cosmological solutions must satisfy it.
Eq.~\p{rfe1} defines a hypersurface in three-dimensional
(${\it \Theta},\theta ,\varpi$)-space, where all orbits of the possible
cosmological solutions lie in.
We call it a constraint hypersurface.
We depict some orbits of cosmological solutions
as well as the fixed points in Fig.~\ref{dynamics_EGBD}.
The arrows denote the evolutionary directions of the orbits.

Solving Eqs.~\p{rfe1}--\p{rfe4}, we find nine fixed points
in Table \ref{t3};
one is the Minkowski spacetime, next four are those of the 4-dimensional
expanding spacetimes, and the rest four are those of contacting spacetimes.

In order to study the stability of those fixed points, we
perturb the variables around the fixed point as
 ${\it \Theta}={\it \Theta}_0+ \delta {\it \Theta}$,
$\theta =\theta_0 + \delta \theta$, and
$\varpi=\varpi_0+\delta \varpi$.
We find the perturbation equations as
\bea
{d\over dt}\left(
\begin{array}{c}
\delta {\it \Theta}\\
 \delta \theta \\
\delta \varpi \\
\end{array}
\right)
= {\cal M}_0
\left(
\begin{array}{c}
\delta {\it \Theta} \\ \delta \theta \\ \delta \varpi
\end{array}
\right),
\ena
where ${\cal M}_0$
is a 3$\times$3 matrix, which is evaluated at the fixed point.
We find one degenerate eigenvalue $M_0=-(3{\it \Theta}_0+6\theta_0-2\varpi_0)$
for the 3$\times$3 matrix ${\cal M}_0$,
 just as the case in the text~\cite{footnote}.
We find that the fixed points of 4-dimensional expanding spacetimes are
always stable,
while those of contracting spacetimes are unstable,
We summarize our result in Table \ref{t3}.

In order to see the deference between the solutions in the string frame
and those in the Einstein frame, we first compare the fixed points.
In the effective theory in the Einstein frame,
they found the eleven fixed points, which values are given in \cite{BGO}.
Although those values are different from our results, the qualitative
properties are very similar to ours as follows:\\
(1) The four-dimensional universe is given by the FLRW spacetime with
power-law expansion.\\
(2) The accelerating universe is found only in the solutions with
a negative power exponent in the limit of $\tau\rightarrow 0$.
In this limit, the string coupling constant $e^\phi$ diverges.
As a result, we are not sure whether such a spacetime is really obtained in
the effective field theory.\\
(3) The expanding universe is stable, while the contracting universe is
unstable.
As for the stability, we find one degenerate eigenvalue for 3$\times$3
perturbation matrix ${\cal M}_0$ just as the case in the text.
On the other hand, for  the Einstein-Gauss-Bonnet-dilaton system
 in the Einstein frame, they found two eigenvalues; one degenerate
value for the metric perturbations and
one for the dilaton field.
It may be because there exists some symmetry between
two metric components ($u_1$ and $u_2$),
 but not for the dilaton field ($\phi$) in the Einstein frame.

We then show the constraint hypersurface in the three dimensional
 $({\it \Theta},\theta,\varpi)$-space.
To compare two cases in detail, we plot several slices with
$\dot \phi = $ constant
($\dot \phi = $0, 0.06992, 0.2133, 0.7, and  2)
 in both frames in Fig.~\ref{hypersurface}.
Since there is the difference in definition of the cosmic times
in the string frame and in the Einstein frame,
we choose $\phi=0$, by which two $\dot \phi$'s become the same.
For $\phi\neq 0$, the values of ${\it \Theta},\theta
$ and $\varpi$ and the slice separations
are stretched, but the topologies of the constraint hypersurfaces do not change.

The cross sections in the string frame and in the Einstein frame
coincide at $\dot \phi = 0$, which is also the same as
the case of the Einstein-Gauss-Bonnet system~\cite{ISHI}.
However the hypersurfaces differ in their topologies on the other slices.
Such a difference may affect the global dynamics of cosmological solutions,
although the local stabilities are the same.

\begin{widetext}

\begin{table}[t]
\begin{center}
\begin{tabular}{c|@{}c@{}|@{}r@{}|c@{}|c|r|@{}r@{}} \hline \hline
 & fixed point (${\it \Theta}_0,\theta_0,\varpi_0$) & $P$~~~~~~~~& ~$H$~ &
 ~A/D~&
$M_0$~~~~ & ~stability~\\ \hline \hline
M & ( 0, 0, 0 ) &$0~$&$0$&-&$0~$& ~marginal\\ \hline
E$_1$ & (~0.689409,$-0.175760,-0.177498$) &
$-0.970990~$~&$+$&A& $ -1.36866~$& stable\\ \hline
E$_2$ & (~0.893673,$-0.112256$, ~0.069923) &
$-1.197415~$~&$+$&A& $-1.86763~$& stable \\ \hline
E$_3$& ($-$0.288681, ~0.321980, ~0.213335) &
$0.616424~$~&$+$&D& $ -0.639169~$& stable\\ \hline
E$_4$& ($-$0.507533, ~0.312010, $-$0.094005) &
$0.507265~$~&$+$&D& $-0.537467~$& stable \\ \hline
C$_1$ &($-$0.689409, ~0.175760, ~0.177498)&
$-0.970990~$&$-$&A&$1.36866~$& unstable  \\ \hline
C$_2$& ($-$0.893673, ~0.112256,$-$0.069923) &
$~~-1.197415~$~&$-$&A& $1.86763~$& unstable \\ \hline
C$_3$& (~0.288681,$-$0.321980,$-$0.213335) &
$0.616424~$~&$-$&D&  $0.639169~$& unstable\\ \hline
C$_4$& (~0.507533, $-$0.312010, ~0.094005) &
$0.507265~$~&$-$&D& $0.537467~$& unstable\\ \hline
\hline
\end{tabular}
\caption{Fixed points of the Einstein-Gauss-Bonnet-dilaton system
in the string frame and their properties. All solutions except for
the Minkowski space (M) show the  power-law expansion as $a=\tau^P$.
The cosmic time in four dimensions $\tau$ takes the value
in the range of $\tau:-\infty \rightarrow 0$ for the negative power exponent
$(P<0)$, while in the range of   $\tau:0\rightarrow \infty $ for the
 positive power
exponent $(P>0)$. $M_0$ is the eigenvalue of the perturbation matrix
 ${\cal M}_0$.
$H$ is the Hubble expansion parameter in four-dimensional our universe.
A and D denote the  accelerating universe and the decelerating universe, respectively.}
\label{t3}
\end{center}
\end{table}

\begin{figure}[h]
\begin{center}
String frame~~~~~~~~~~~~~~~~~~~~~~~~~~~Einstein frame\\[1em]
 \includegraphics[width=4cm,height=2.4cm,clip]{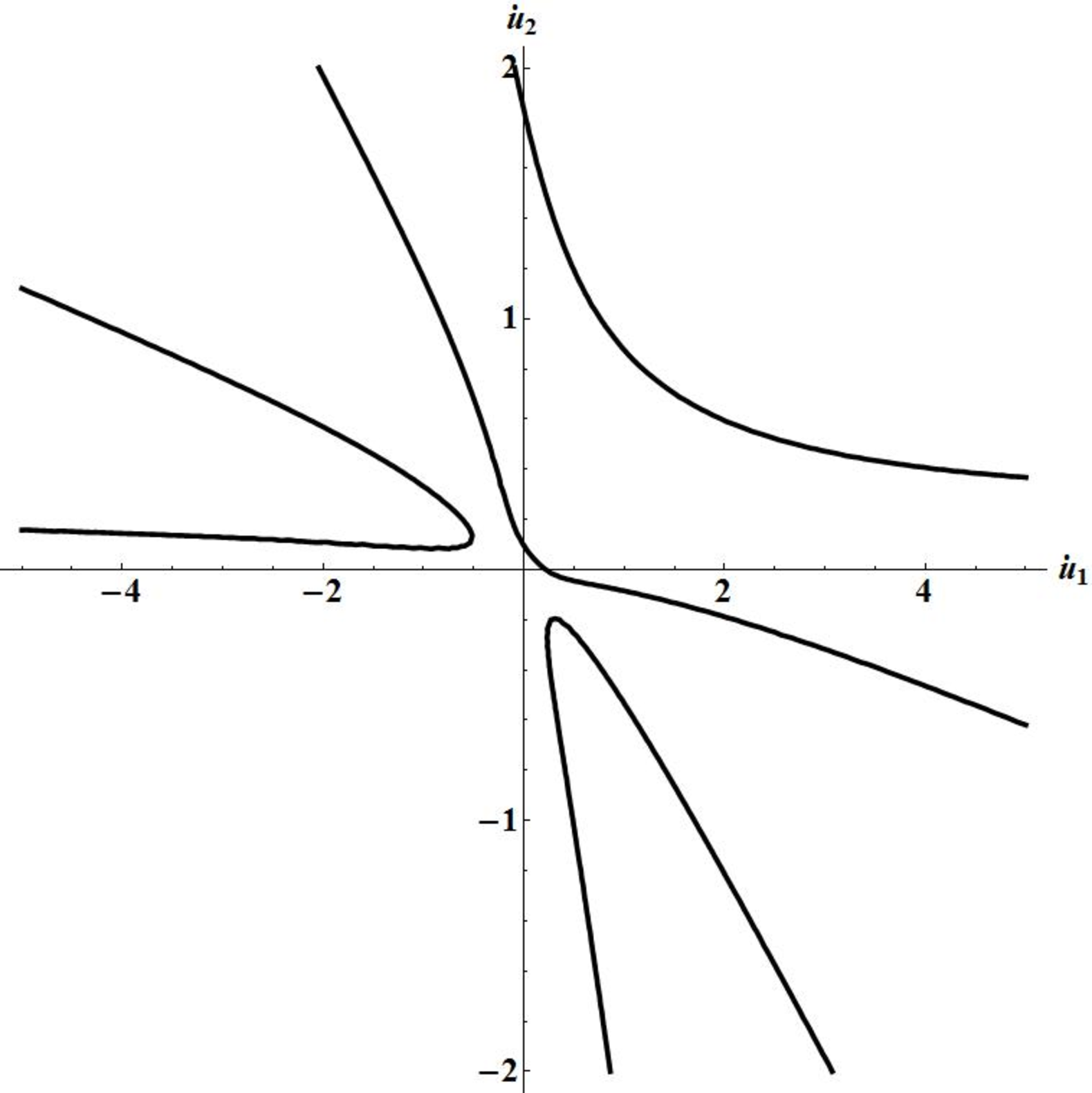} \qquad
 \includegraphics[width=4cm,height=2.4cm,clip]{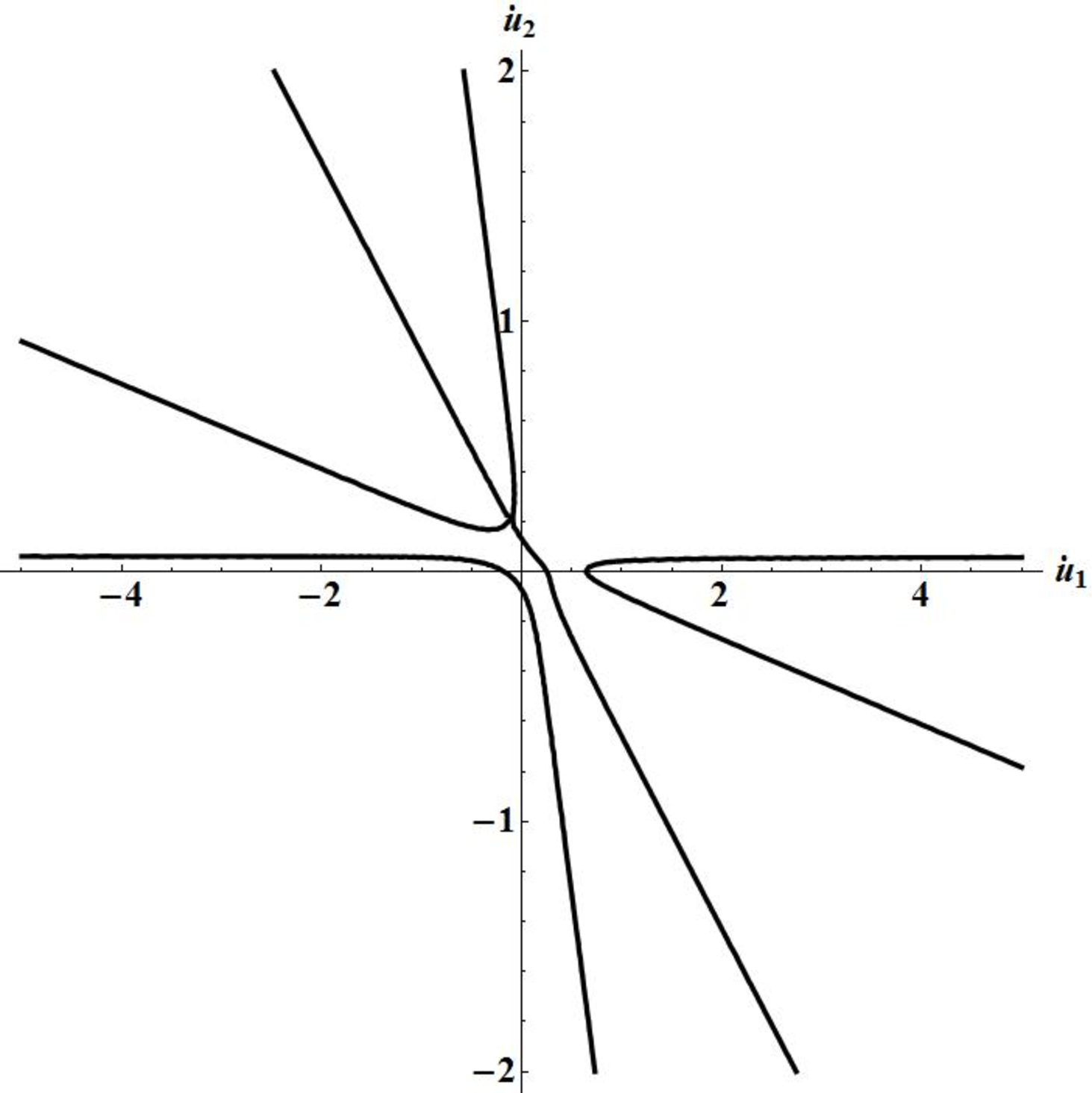}\\
slice at $\dot \phi$=0.7~~~~~~~~~~~~~~~~~~~~~~~slice at $\dot \phi$=0.7\\[.4em]
 \includegraphics[width=4cm,height=2.4cm,clip]{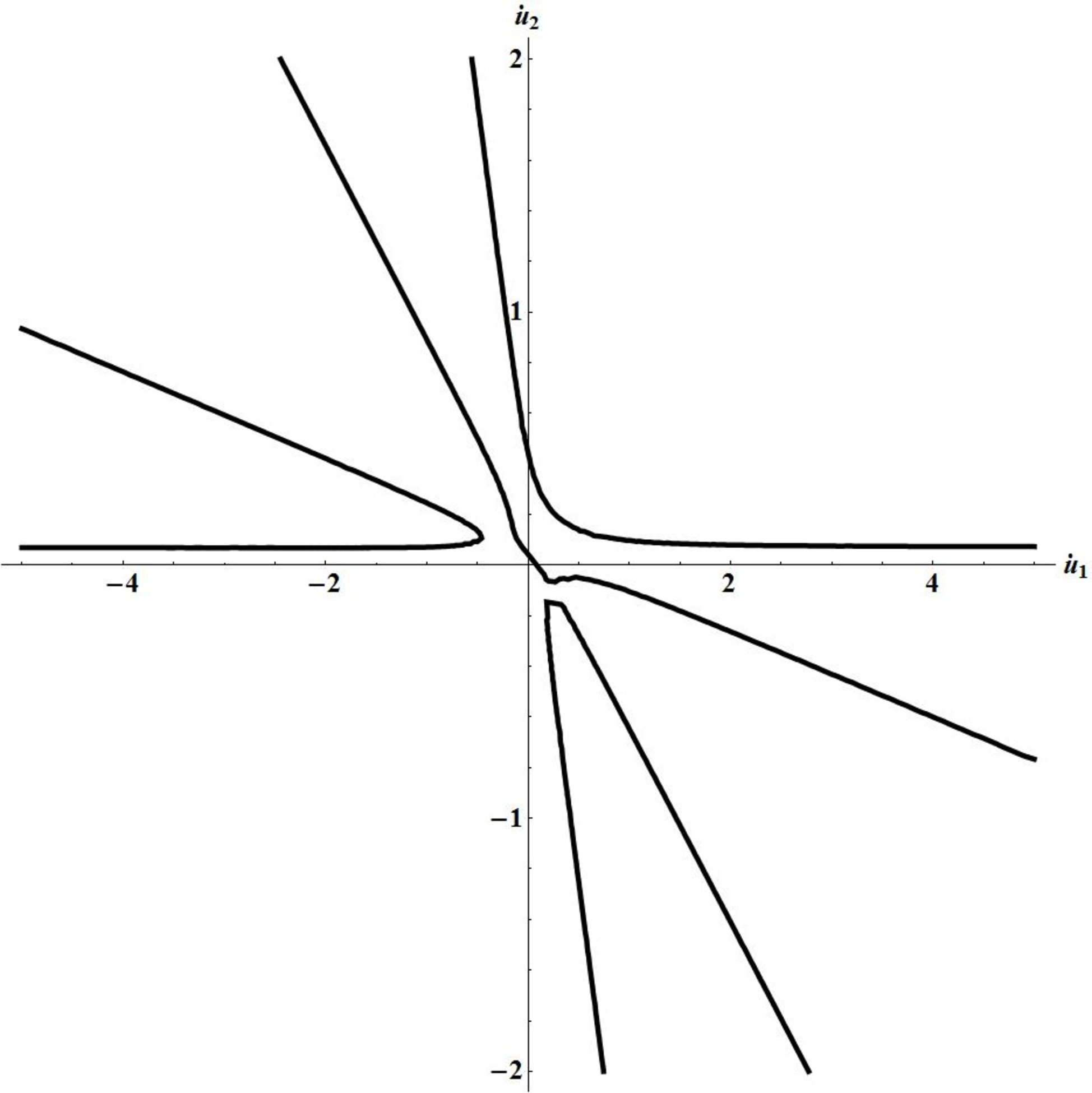} \qquad
 \includegraphics[width=4cm,height=2.4cm,clip]{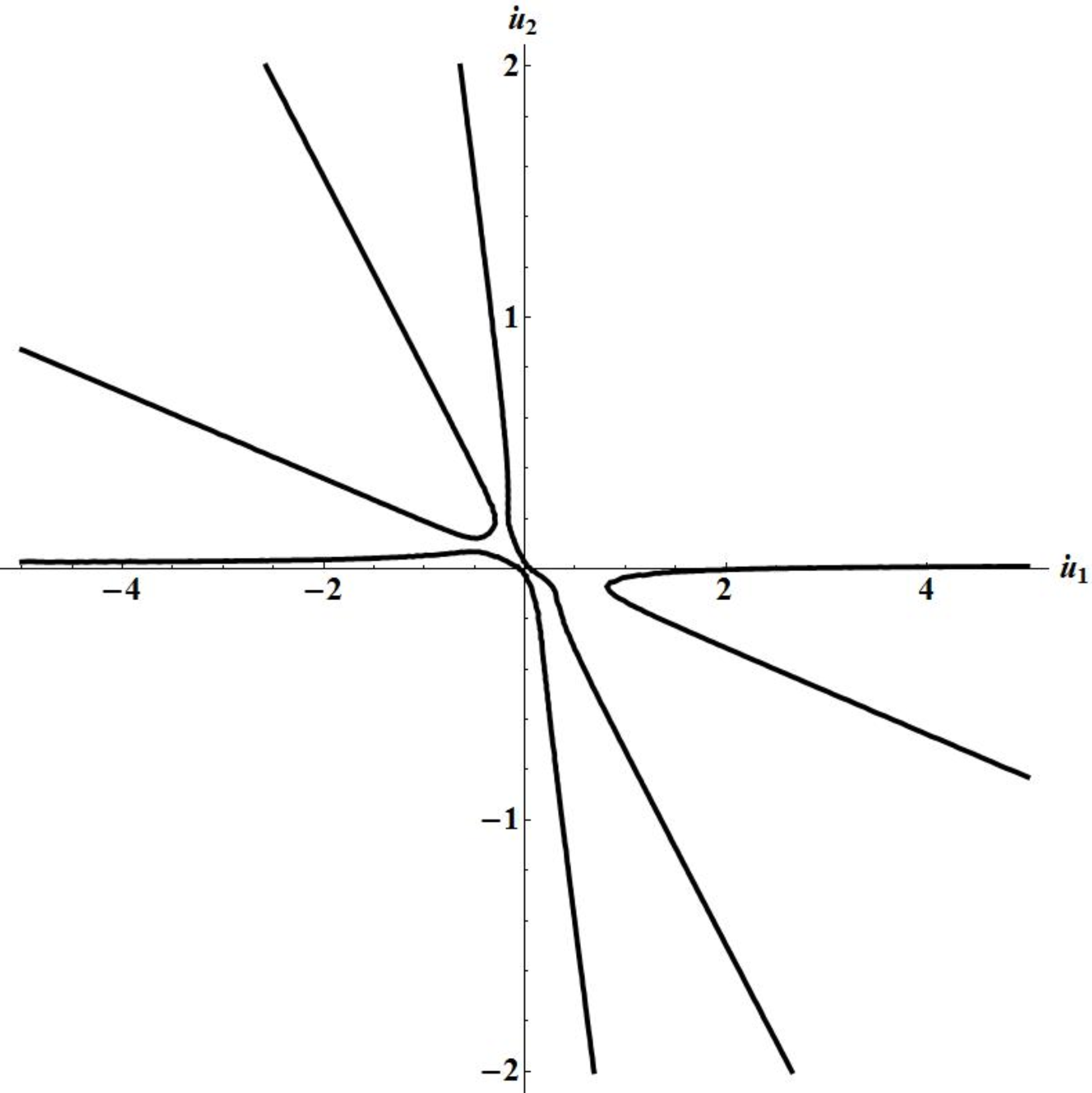}\\
slice at $\dot \phi$=0.2133,~~~~~~~~~~~~~~~~~~~~~~~slice at $\dot \phi$=0.2133
\\[.4em]
 \includegraphics[width=4cm,height=2.4cm,clip]{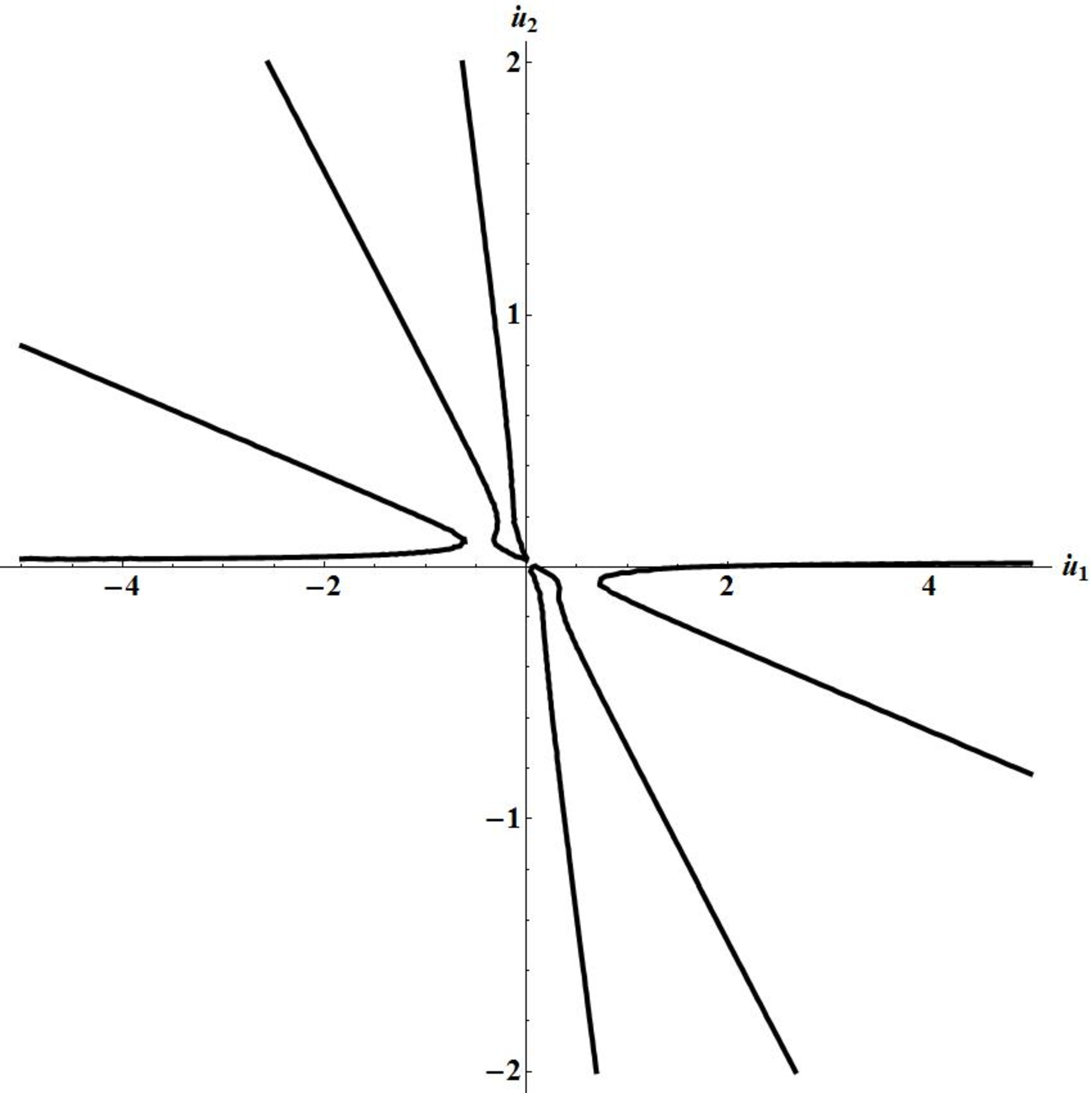} \qquad
 \includegraphics[width=4cm,height=2.4cm,clip]{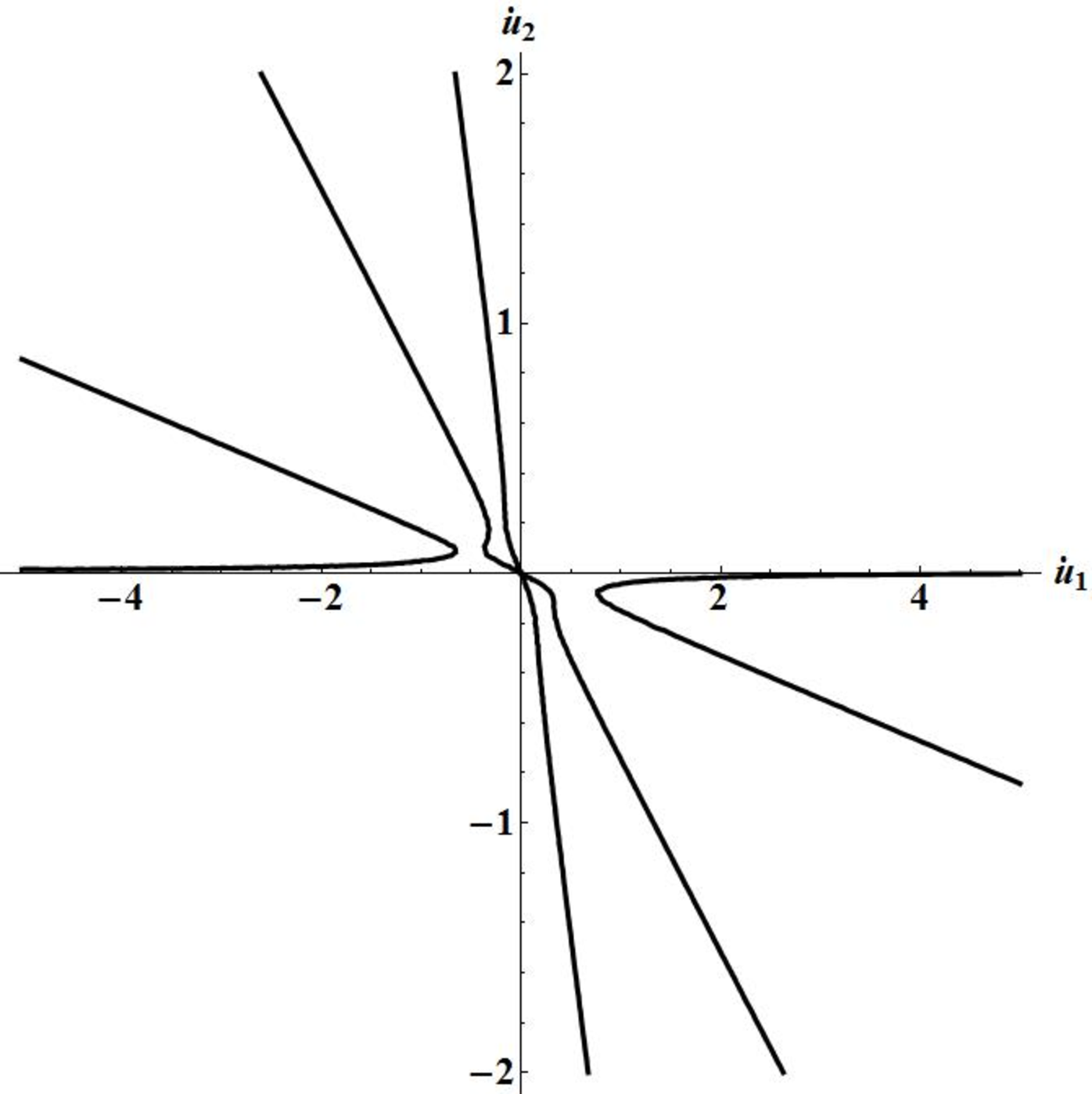}
 \\ slice at $\dot \phi$=0.06992
~~~~~~~~~~~~~~~~~~~~~~~slice at $\dot \phi$=0.06992
\\[.4em]
 \includegraphics[width=4cm,height=2.4cm,clip]{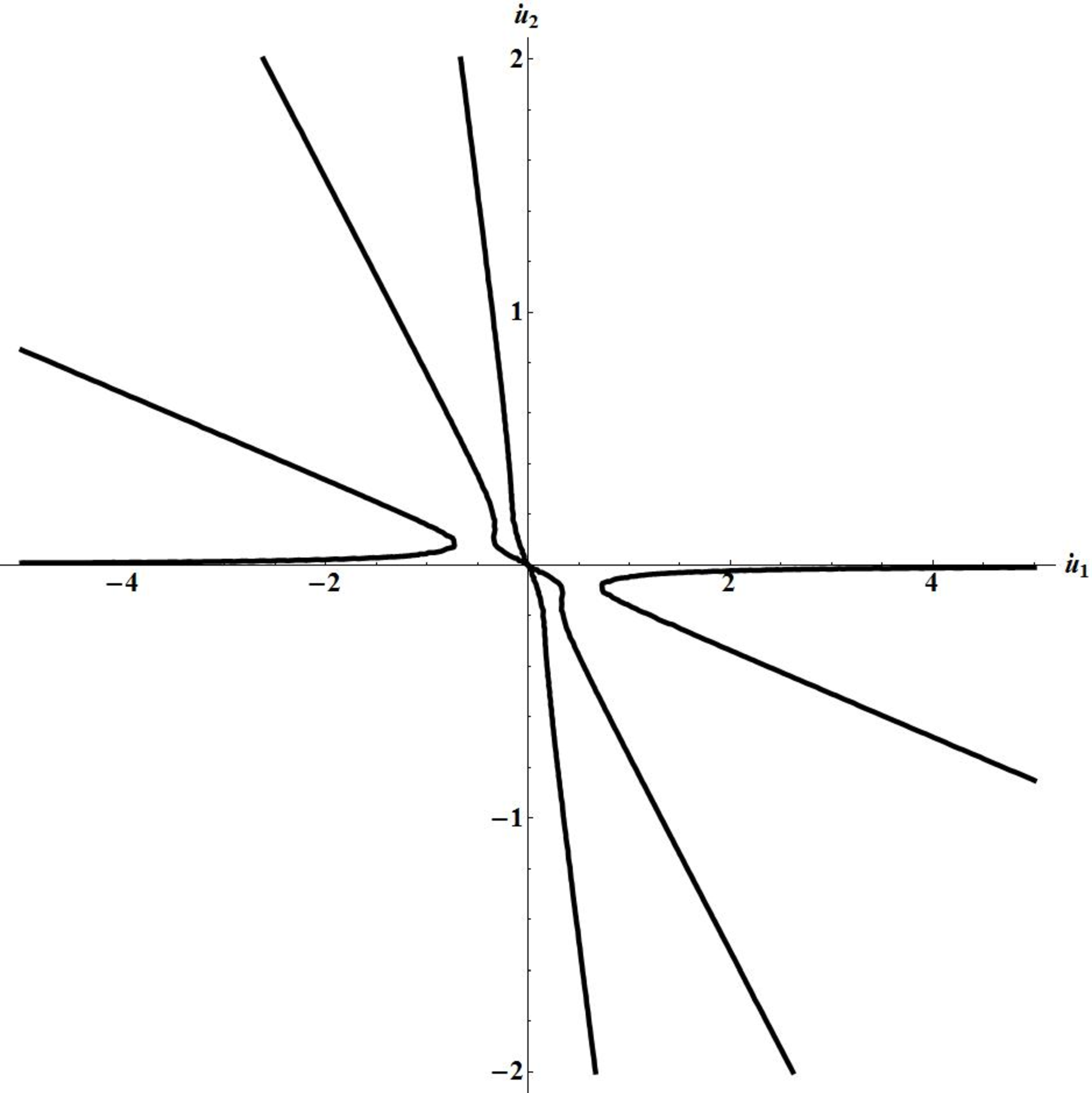}\qquad
 \includegraphics[width=4cm,height=2.4cm,clip]{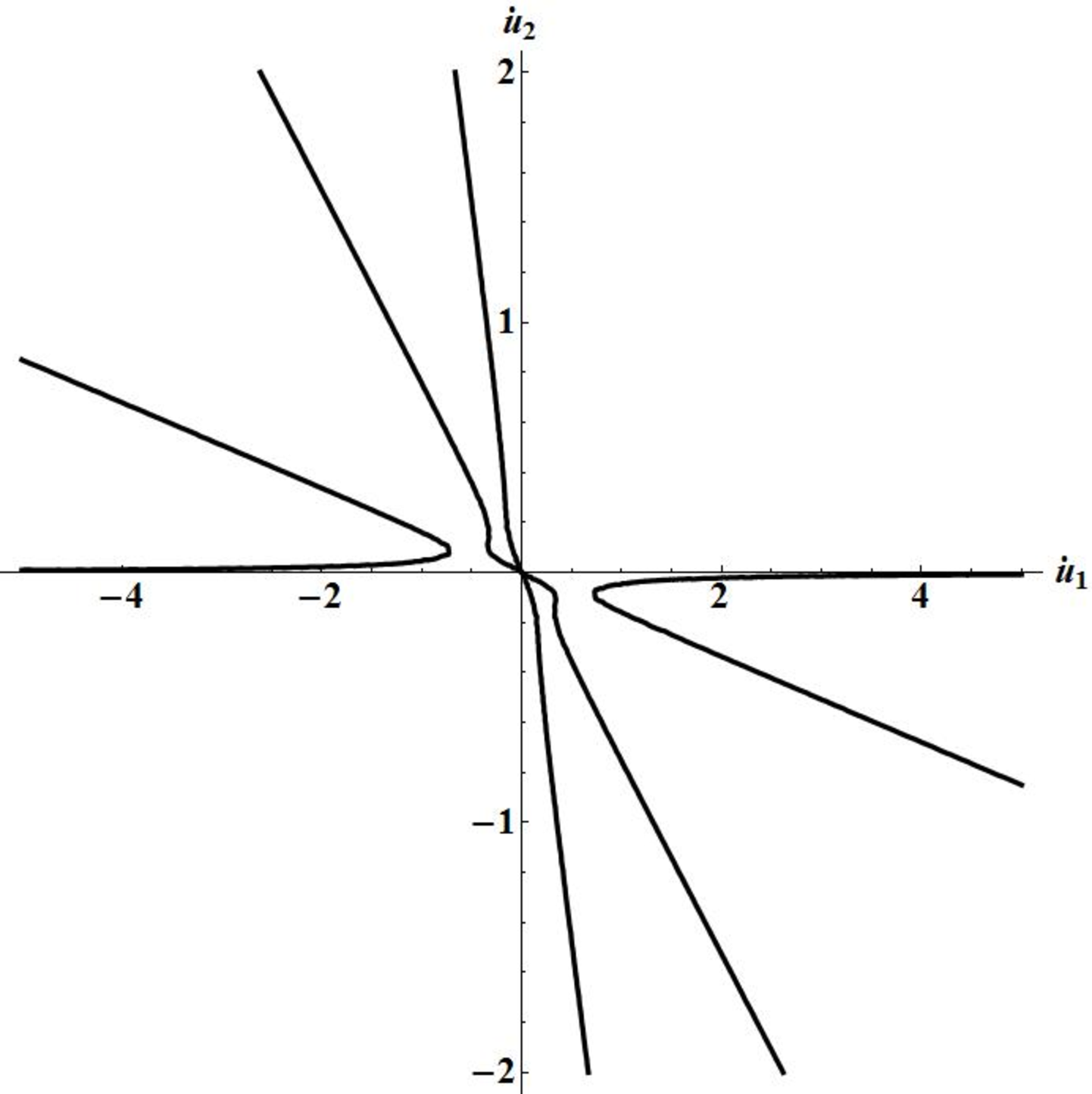}\\
slice at $\dot \phi$=0~~~~~~~~~~~~~~~~~~~~~~~slice at $\dot \phi$=0\\
\end{center}
\caption{The $\dot \phi$=constant slices of the constraint hypersurfaces
 which satisfy the constraint equation
(\ref{rfe1}) in the effective theory in the string frame
and that in the Einstein frame.
We set $\phi=0$ to find the same cosmic time in both frames.
Although the cross sections coincide on the slice
of $\dot \phi=0$, the hypersurfaces differ their topologies
on the other slices. }
\label{hypersurface}
\end{figure}

\end{widetext}



\begin{thebibliography}{99}
\bibitem{wmap}
D.~N.~Spergel {\it et al.}  [WMAP Collaboration],
  Astrophys.\ J.\ Suppl.\  {\bf 148}, 175 (2003)
  [arXiv:astro-ph/0302209];
  Astrophys.\ J.\ Suppl.\  {\bf 170}, 377 (2007)
  [arXiv:astro-ph/0603449];\\
H.~V.~Peiris {\it et al.}  [WMAP Collaboration],
  Astrophys.\ J.\ Suppl.\  {\bf 148}, 213 (2003)
  [arXiv:astro-ph/0302225].

\bibitem{Starobinsky}
A.~A.~Starobinsky,
   Phys.\ Lett.\ B {\bf 91}, 99 (1980).

\bibitem{inflation1}
K. Sato, Mon. Not. Roy. Astron. Soc.{\bf 195}, 467 (1981);\\
A. H. Guth, Phys. Rev. D {\bf 23}, 347 (1981).

\bibitem{inflation2}
A. Albrecht and P.J. Steinhardt, Phys. Rev. Lett. {\bf 48}, 1220 (1982);\\
A. D. Linde, Phys. Lett. B {\bf 108}, 389 (1982).

\bibitem{inflation3}
A. D. Linde, Phys. Lett B {\bf 129}, 177 (1983).

\bibitem{inflation4}
See also the following review articles:
A. D.~Linde,
  [arXiv:hep-th/0503203v1];\\
A. D.~Linde,
J. Phys.: Conf. Ser. {\bf 24}, 151
[arXiv:hep-th/0503195];\\
A. D.~Linde,
Lect. Notes Phys. {\bf 738}, 1 (2008)
 [arXiv:0705.0164 [hep-th]] ;\\
L.~McAllister and E.~Silverstein,
Gen. Rel. Grav. {\bf 40}, 565 (2008)
[arXiv:0710.2951 [hep-th]];\\
D.~H.~Lyth,
Lect. Notes Phys. {\bf 738}, 81 (2008)
[arXiv: hep-th/0702128]

\bibitem{no-go}
G.~W.~Gibbons,
 {\it Proceedings of the GIFT Seminar on Theoretical Physics, San Feliu de
 Guixols, Spain, Jun 4-11, 1984},
 ed. F.~Del~Aguila, {\it et al.} (World Scientific, 1984) pp.~123-146;\\
J.~M.~Maldacena and C.~Nunez,
 Int.\ J.\ Mod.\ Phys.\ A {\bf 16}, 822 (2001)
 [arXiv:hep-th/0007018].

\bibitem{TW}
P.~K.~Townsend and M.~N.~R.~Wohlfarth,
 Phys.\ Rev.\ Lett.\  {\bf 91}, 061302 (2003)
 [arXiv:hep-th/0303097].

\bibitem{NO}
N.~Ohta,
 Phys.\ Rev.\ Lett.\  {\bf 91}, 061303 (2003)
 [arXiv:hep-th/0303238];
 Prog.\ Theor.\ Phys.\  {\bf 110}, 269 (2003)
 [arXiv:hep-th/0304172].

\bibitem{Wohlfarth:2003ni}
M.~N.~R.~Wohlfarth,
  Phys.\ Lett.\ B {\bf 563}, 1 (2003)
  [arXiv:hep-th/0304089].

\bibitem{Sbrane1}
C.~M.~Chen, D.~V.~Gal'tsov and M.~Gutperle,
  Phys.\ Rev.\ D {\bf 66}, 024043 (2002)
  [arXiv:hep-th/0204071];\\
N.~Ohta,
  Phys.\ Lett.\ B {\bf 558}, 213 (2003)
  [arXiv:hep-th/0301095].

\bibitem{Sbrane2}
M.~Kruczenski, R.~C.~Myers and A.~W.~Peet,
  JHEP {\bf 0205}, 039 (2002)
  [arXiv:hep-th/0204144];\\
V.~D.~Ivashchuk,
  Class.\ Quant.\ Grav.\  {\bf 20}, 261 (2003)
  [arXiv:hep-th/0208101];\\
see also H.~Lu, S.~Mukherji, C.~N.~Pope and K.~W.~Xu,
  Phys.\ Rev.\ D {\bf 55}, 7926 (1997)
  [arXiv:hep-th/9610107].

\bibitem{Sbrane3}
L.~Cornalba and M.~S.~Costa,
  Phys.\ Rev.\ D {\bf 66}, 066001 (2002)
  [arXiv:hep-th/0203031];\\
S.~Roy,
  Phys.\ Lett.\ B {\bf 567}, 322 (2003)
  [arXiv:hep-th/0304084];\\
A.~Buchel and J.~Walcher,
  JHEP {\bf 0305}, 069 (2003)
  [arXiv:hep-th/0305055];\\
C.~Armendariz-Picon and V.~Duvvuri,
  Class.\ Quant.\ Grav.\  {\bf 21}, 2011 (2004)
  [arXiv:hep-th/0305237];\\
C.~P.~Burgess, P.~Martineau, F.~Quevedo, G.~Tasinato and I.~Zavala C.,
  JHEP {\bf 0303}, 050 (2003)
  [arXiv:hep-th/0301122];\\
I.~P.~Neupane and D.~L.~Wiltshire,
  Phys.\ Lett.\ B {\bf 619}, 201 (2005)
  [arXiv:hep-th/0502003];
I.~P.~Neupane and D.~L.~Wiltshire,
  Phys.\ Rev.\ D {\bf 72}, 083509 (2005)
  [arXiv:hep-th/0504135].

\bibitem{other}
L.~Cornalba and M.~S.~Costa,
  Fortsch.\ Phys.\  {\bf 52}, 145 (2004)
  [arXiv:hep-th/0310099];\\
V.~Balasubramanian,
  Class.\ Quant.\ Grav.\  {\bf 21}, S1337 (2004)
  [arXiv:hep-th/0404075];\\
N.~Ohta,
  Int.\ J.\ Mod.\ Phys.\ A {\bf 20}, 1 (2005)
  [arXiv:hep-th/0411230].

\bibitem{cosm2}
R.~Emparan and J.~Garriga,
JHEP {\bf 0305}, 028 (2003)
[arXiv:hep-th/0304124].

\bibitem{cosm3}
C.~M.~Chen, P.~M.~Ho, I.~P.~Neupane, N.~Ohta and J.~E.~Wang,
JHEP {\bf 0310}, 058 (2003)
[arXiv:hep-th/0306291];
  JHEP {\bf 0611}, 044 (2006)
  [hep-th/0609043].

\bibitem{brane_inflation}
G.R.~Dvali and S.-H.H.~Tye,
  Phys.\ Lett.\ B {\bf 450}, 72 (1999)
  [arXiv;hep-th/9812483];\\
S.B.~Giddings, S.~Kachru and J.~Polchinski,
  Phys.\ Rev.\ D {\bf 66}, 106006 (2002)
  [arXiv:hep-th/0105097];\\
S.~Kachru, R.~Kallosh, A.~Linde, and S.P.~Trivedi,
  Phys.\ Rev.\ D {\bf 68}, 046005 (2003)
  [arXiv:hep-th/0301240];\\
S.~Kachru, R.~Kallosh, A.~Linde, J.~Maldacena, L.~McAllister and S.P.~Trivedi,
 JCAP {\bf 0310} (2003) 013,
  [arXiv:hep-th/0308055].

\bibitem{Brane_rev}
See also the following review article:
S.-H.H.~Tye
 Lect. Notes Phys. {\bf 737}, 949 (2008)
 [arXiv:hep-th/0610221v2].

\bibitem{MT}
  R.~R.~Metsaev and A.~A.~Tseytlin,
  Nucl.\ Phys.\  B {\bf 293}, 385 (1987).

\bibitem{hetero0}
M.~de Roo, H.~Suelmann and A.~Wiedemann,
  Nucl.\ Phys.\ B {\bf 405}, 326 (1993)
  [arXiv:hep-th/9210099].

\bibitem{hetero}
A.~A.~Tseytlin,
  Nucl.\ Phys.\ B {\bf 467}, 383 (1996)
  [arXiv:hep-th/9512081].

\bibitem{Mth}
K.~Peeters, P.~Vanhove and A.~Westerberg,
  Class.\ Quant.\ Grav.\  {\bf 18}, 843 (2001)
  [arXiv:hep-th/0010167].

\bibitem{TBB}
A.~A.~Tseytlin,
  Nucl.\ Phys.\ B {\bf 584}, 233 (2000)
  [arXiv:hep-th/0005072];\\
K.~Becker and M.~Becker,
  JHEP {\bf 0107}, 038 (2001)
  [arXiv:hep-th/0107044].

\bibitem{maeda1}
K. Maeda, Phys. Rev. D {\bf 37}, 858 (1988).

\bibitem{KK_cosmology}
See, e.g.
  ``Modern Kaluza-Klein Theories,''
  ed. T.~Appelquist, A.~Chodos and P.~G.~O.~Freund (1987, Addison-Wesley),
  Chap. VI.

\bibitem{maeda}
K.~Maeda,
Class. Quant. Grav. {\bf 3}, 233 (1986);
Class. Quant. Grav. {\bf 3}, 651 (1986);
K. Maeda  and H. Nishino, Phys. Lett. B {\bf 154}, 358 (1985);
Phys. Lett. B {\bf 158}, 381 (1985).

\bibitem{ISHI}
H.~Ishihara,
  Phys.\ Lett.\ B {\bf 179}, 217 (1986).

\bibitem{old1}
K.~Maeda,
   Phys.\ Lett.\ B {\bf 166}, 59 (1986);\\
J.~R.~Ellis, N.~Kaloper, K.~A.~Olive and J.~Yokoyama,
   Phys.\ Rev.\ D {\bf 59}, 103503 (1999)
   [arXiv:hep-ph/9807482].

\bibitem{MO}
K.~Maeda and N.~Ohta,
Phys.\ Lett.\  B {\bf 597}, 400 (2004)
[arXiv:hep-th/0405205];
K.~Maeda and N.~Ohta,
Phys.\ Rev.\  D {\bf 71}, 063520 (2005)
[arXiv:hep-th/0411093];\\
K.~Akune, K.~Maeda and N.~Ohta,
Phys.\ Rev.\  D {\bf 73}, 103506 (2006)
[arXiv:hep-th/0602242].

\bibitem{BGO}
K.~Bamba, Z.~K.~Guo and N.~Ohta,
  Prog.\ Theor.\ Phys.\  {\bf 118}, 879 (2007)
  [arXiv:0707.4334 [hep-th]].

\bibitem{Sahdev}
D. Sahdev, Phys. Lett. B {\bf 137}, 155 (1984);\\
F. Lucchin and S. Matarrese, Phys. Rev. D{\bf 32}, 1316  (1985).

\bibitem{GMQS}
S.~R.~Green, E.~J.~Martinec, C.~Quigley and S.~Sethi,
  arXiv:1110.0545 [hep-th].

\bibitem{GOT}
Z.~K.~Guo, N.~Ohta and S.~Tsujikawa,
  Phys.\ Rev.\  D {\bf 75}, 023520 (2007)
  [arXiv:hep-th/0610336].

\bibitem{GS}
  Z.~-K.~Guo and D.~J.~Schwarz,
  Phys.\ Rev.\  D {\bf 80}, 063523 (2009)
  [arXiv:0907.0427 [hep-th]].


\bibitem{conf}
M.~Dabrowski, J.~Garecki and D.~Blaschke,
Ann.\ Phys.\ {\bf 18}, 13 (2009)
[arXiv:0806.2683[gr-qc]];\\
K.~Maeda, N.~Ohta and Y.~Sasagawa,
Phys.\ Rev.\ D {\bf 80}, 104032 (2009)
[arXiv:0908.4151[hep-th]];\\
M.~Iihoshi,
  Gen.\ Rel.\ Grav.\  {\bf 43}, 1571 (2011)
  [arXiv:1011.2088 [hep-th]].

\bibitem{footnote1}
We would like to take this opportunity to correct a typo in our paper~\cite{conf}:
The first term in the last line of Eq.~(2.5) should be
$D(D-3) \gamma^2 G_{\mu\nu}\nabla^\mu \phi\nabla^\nu \phi$
instead of $D(D-3) \gamma^2 G_{\mu\nu}\nabla^\mu \nabla^\nu \phi$.

\bibitem{galileon}
A. Nicolis, R. Rattazzi and E. Trincherini,
Phys. Rev. D {\bf 79}, 064036 (2009);\\
C. Deffayet, G. Esposito-Farese and A. Vikman,
Phys. Rev. D {\bf 79}, 084003 (2009);\\
C. Deffayet, S. Deser and
G. Esposito-Farese, Phys. Rev. D {\bf 80}, 064015 (2009);\\
C. Deffayet, O. Pujolas, I. Sawicki and A. Vikman,
JCAP {\bf 10}, 026 (2010);\\
T.~Kobayashi, M.~Yamaguchi, and J.~Yokoyama
Phys.\ Rev.\ Lett. {\bf 105}, 231302 (2010)
[arXiv:1008.0603 [hep-th]];\\
T.~Kobayashi, M.~Yamaguchi, and J.~Yokoyama
Prog.\ Theor.\ Phys. {\bf 126}, 511 (2011)
[arXiv:1105.5723 [hep-th]].

\bibitem{foonote2}
We can confirm from Eq. (\ref{Bianchi})
that it gives the fixed points if $\dot \phi\neq 0$.
However it is not the case if $\dot \phi = 0$.
(See Appendix \ref{AppA3}.)

\bibitem{footnote}
We have performed stability analysis for arbitrary values of $\mu$,
 $\nu$ and $\lambda$ in order to include the case of the
Einstein-Gauss-Bonnet-delaton system in the string frame ($\mu=\nu=\lambda=0$).
We find that three eigenvalues are always degenerate and its value $M_0$ does not
depend on those coupling constants $\mu, \nu,$and $\lambda$,
but is given by the fixed point $({\it \Theta}_0, \theta_0,\varpi_0)$
as $M_0=-(3{\it \Theta}_0+6\theta_0-2\varpi_0)$.

\bibitem{conformal_transformation}
G. Magnano, M. Ferraris and M. Francaviglia,
Gen. Rel. Grav. {\bf 19}, 465 (1987);\\
A. Jakubiec and J. Kijowski, Phys. Rev. D {\bf 37}, 1406 (1988);\\
K. Maeda, Phys. Rev. D {\bf 39}, 3159 (1989).

\end{thebibliography}
\end{document}